\documentclass[a4paper,10pt]{article}
\usepackage{jheppub} 

\usepackage[pagewise]{lineno}

\usepackage[sorting=none,style=numeric-comp]{biblatex}
\usepackage{graphicx}

\usepackage{amsmath}
\usepackage{amsfonts}
\usepackage{amssymb}
\usepackage{mathtools}
\usepackage{mathrsfs}

\usepackage{xcolor}

\newcommand{\lie}[1]{\mathfrak{#1}}
\newcommand{\killing}[2]{\left\langle {#1} , {#2} \right\rangle}
\newcommand{\res}[1]{\text{Res}_{#1} \ } 
\newcommand{\tr}[3]{ \langle \! \langle {#1} , {#2} \rangle \! \rangle_{({#3})} } 
\newcommand{\bigtr}[3]{ \bigl\langle \negmedspace \bigl\langle {#1} , {#2} \bigr\rangle \negmedspace  \bigr\rangle_{({#3})} } 
\newcommand{\Bigtr}[3]{ \Bigl\langle \hspace{-1.25mm} \Bigl\langle {#1} , {#2} \Bigr\rangle \hspace{-1.25mm}  \Bigr\rangle_{({#3})} }

\newcommand{\ca}{\, ,}
\newcommand{\fp}{\, .} 

\usepackage{cleveref}

\title{\boldmath Hamiltonian Analysis of Doubled 4d Chern-Simons} 

\author{Jake Stedman}
\affiliation{King's College London,\\
Strand, 
London, 
WC2R 2LS}

\emailAdd{jake.williams@kcl.ac.uk}

\abstract{
Motivated by a conjecture that doubled four-dimensional Chern-Simons produces new integrable models, we perform its Hamiltonian analysis and find the theory's Poisson algebra. 
This requires carefully accounting for a set of boundary conditions that identify two gauge fields. 
Two methods for doing so are given, one of which is based on edge-modes and the other on a recharacterisation of the boundary conditions as constraints. 
We find that the Poisson algebra is that of an affine Gaudin model subject to a constraint, generalising the Goddard-Kent-Olive construction (from conformal field theory) to the world of integrable models. We also conjecture the existence of extended quantum groups 
and a generalisation of the affine Harish-Chandra Isomorphism.
}

\begin{document}
\maketitle
\flushbottom

\section{Introduction \label{sec: intro}} 

This paper is a continuation of our previous work in \cite{Stedman:2021wrw}, in which we gave a method for constructing gauged $\sigma$-models from two coupled four-dimensional Chern-Simons (4dCS) theories \cite{Costello:2013zra,Costello:2013sla,Costello:2017dso,Costello:2018gyb,Costello:2019tri}, called doubled 4dCS. 
Due to the nature of the construction, the resulting models have a crucial property that, in a single 4dCS theory, is a hallmark of integrability. 
This is that their equations of motion are given by two Lax equations, \textit{i.e.}, they are each the flatness condition for a two-dimensional connection with a (potentially trivial) meromorphic dependence upon an auxiliary complex parameter \cite{Lax:1968fm,Zakharov:1979zz}. 
For this reason, it was conjectured that the models were integrable. 
This paper constitutes the first step in proving this. 

Generally a classical field theory is said to be integrable if there exists an infinite set of Poisson-commuting charges. 
Usually these charges can be constructed from a theory's Lax connection\cite{Olive:1984mb,Lacroix:2017isl,Babelon:2003qtg}, making its existence particularly important. 
Naively, we might hope that the task of proving integrability within these new models is relatively simple. 
Surely all you need to do is follow the standard methods for constructing charges from a Lax connection? 
As we will see, this guess is incorrect. 

To prove integrability we need to: $i)$ identify the Poisson structure of the Lax connections, and $ii)$ find the set of commuting charges. 
The purpose of this paper is to perform the Hamiltonian analysis of doubled 4dCS, and thus achieve step $i)$. 
Our principal result is the following: given a finite set of points $P_N \subset \mathbb{C} P^1$, we find for each $q \in P_N$ a system of pairs $(\Gamma_q , \gamma_q) \in \lie g \times \mathcal G$, where $\mathcal G$ is a Lie group and $\lie g$ its associated algebra. 
Each of these pairs satisfies the non-trivial Poisson brackets: 
\begin{equation}
    \{ \Gamma_{\! q \textbf{1}}(x) , \Gamma_{\! q \textbf{2}}(y) \} = [\mathsf{C} , \Gamma_{\! q \textbf{1}}(x)] \delta(x-y) - \frac{k_q}{2 \pi} \mathsf{C} \delta'(x-y) \ca \quad\{ \Gamma_{\! q \textbf{1}} (x), \gamma_{q \textbf{2}} (y) \} = \gamma_{q \textbf{2}} (y)  \mathsf C \delta (x-y) \ca
\end{equation} 
in which $\mathsf C$ is the split Casimir of $\lie g$ (defined below), and they are collectively subject to the `first-class constraint': 
\begin{equation}
    \sum_{q \in P_N} \left( \Gamma_q (x) + \frac{k_q}{2 \pi} \gamma_q' (x) \gamma_q^{-1} (x) - \gamma_q (x) \Gamma_q (x) \gamma_q^{-1} (x) \right)^{\lie h} \approx 0 \ca \label{eq:remaining-constraint}
\end{equation} 
where the superscript $\lie h$ denotes the projection into a subalgebra $\lie h$ of $\lie g$. 
The reader can think of this as an extension of the Goddard-Kent-Olive (GKO) construction from conformal field theory \cite{Goddard:1984vk,Goddard:1986ee,Bais:1987zk} to the world of integrable models. 
We thus expect to find many new integrable theories characterised by the above equations. 
Step $ii)$ requires additional work and will thus be done elsewhere. 
We will explain why this is the case in section \ref{section:coset-ifts}. 

The coupling between the two 4dCS theories (defined in the next section) is special as it is localised to certain two-dimensional defects that each sit at points in the Riemann sphere, while spanning a cylinder of radius $R$. 
The inclusion of these terms has the effect of modifying a set of boundary conditions such that some of the components of the theory's two gauge fields $A$ and $B$ are equal. 
This identification was easily dealt with in the Lagrangian analysis performed in \cite{Stedman:2021wrw}, as all we needed to do was account for it when solving a series of differential equations. 
The same is not true of Hamiltonian analysis, in which a new approach is necessary. 
We describe two such techniques. 
In the first, the identification can be thought of as gauging improper gauge transformations to render them proper. 
This interpretation lends itself nicely to an analysis of edge modes given in section \ref{sec:edgemodes}. 
The second approach recasts the boundary conditions as (first-class) constraints to which we can then apply Dirac's algorithm \cite{Dirac-1964}. 
Similar ideas to this latter approach have appeared elsewhere, see \textit{e.g.} \cite{Corichi:2020wdp}, although our justifications for doing this differ\footnote{In \cite{Corichi:2020wdp} authors treat the boundary conditions as constraints from the very beginning, whereas we only do so after eliminating a set of bulk variables.}. 
We expect this recasting of boundary conditions as constraints to be applicable more generally when there exists a gauge symmetry at a boundary/on defects. 

The structure of this paper is as follows. 
In section \ref{sec: Action}, we introduce the doubled 4dCS action and discuss the notion of functional differentiability introduced by Regge and Teitelboim in \cite{Regge:1974zd}. 
This latter concept is particularly important in Hamiltonian analysis, as it ensures that the Poisson brackets between certain functionals are well-defined. 
We conclude the section with an explanation of why the doubled 4dCS gauge group is of a particular form. 

In section \ref{sec: Bulk Hamiltonian analysis}, we perform the `bulk' Hamiltonian analysis of doubled 4dCS by following Dirac's algorithm. 
This starts with a derivation of the theory's Hamiltonian, Poisson brackets and `primary constraints'. 
As usual, these constraints can be divided into two subsets, called `first' and `second-class', enabling us to perform two actions. 
The first action is the derivation of a set of `secondary constraints' from the requirement that the members of the former class hold for all time. 
The second action is the elimination of the latter class via a projection of the Poisson brackets. 
These two steps are then repeated for the secondary constraints until no others are found. 
We finish the section with a discussion of edge-modes that can be skipped on first reading. 

The first-class constraints of section \ref{sec: Bulk Hamiltonian analysis} are then fixed in section \ref{sec: Bulk Gauge fixing} by imposing a series of gauge choices. 
Doing this enables us to eliminate the constraints and find a new set of Poisson brackets. 

In the penultimate section (\ref{section:boundary-Hamiltonian-Analysis}), we argue that the boundary conditions can be recast as constraints. 
Once armed with this viewpoint, we perform Dirac's algorithm to find that these `boundary constraints' are first-class. 
Unlike with their bulk cousins, we can eliminate all but one of the boundary constraints by introducing and working with a set of gauge-invariant observables. 
After doing this, we are left with \cref{eq:remaining-constraint} as the remaining constraint and finish the section with a basic definition of the integrable models defined by it.  
This is done by making an analogy with the GKO construction. 

We conclude by discussing several potential future directions for this research in section \ref{sec:future-directions}. 

\acknowledgments

We would like to thank Beth Romano, Sakura Schafer-Nameki, Joe Smith, and G{\'e}rard Watts for helpful discussions during the course of this work. 
We also wish to thank the Isaac Newton Institute for Mathematical Sciences and the organisers of the programme \textit{Quantum field theory with boundaries, impurities, and defects} at which some of this research was performed. 
This work was partially funded by the STFC grant ST/T506187/1. 

\section{The Action, Functional Differentiability and Gauge Symmetry Group \label{sec: Action}} 

This section is split into two. 
In the first part, we define the basic structures from which the doubled 4dCS action is constructed. 
This includes a meromorphic one-form $\omega$, and a pair of connections $(A,B)$ on a principal bundle $\mathcal{P}$. 
Having done this, we define the doubled action. 
In the second subsection, we introduce the notion of functional differentiability and describe the role of boundary terms within this concept. 
Finally, we conclude by explaining why the gauge group of the theory is of a particular form, which will be given shortly. 

\subsection{Defining the Action \label{subsec: action and bundle}}

To define the doubled 4dCS action we need two objects: a one-form $\omega$ and a principal bundle $\mathcal{P}$. 
The former is defined as follows. 
Given a copy of the Riemann sphere we take $\omega \in \Omega^{(1,0)}(\mathbb{C} P^1)$ to be meromorphic with a finite number of poles, $N_P$, and zeros, $N_Z$, where $N_P = N_Z +2$ by the Riemann-Roch theorem. 
The poles and zeros of $\omega$ are, for simplicity, assumed to be simple/first order, and we denote their respective sets by $P$ and $Z$. 
To parametrise the Riemann sphere, we choose two antipodal points, $p$ and $q$, and cover it with the charts $N = \mathbb{C} P^1 \setminus p$ and $S = \mathbb{C} P^1 \setminus q$. 
The complex coordinates of these charts are $(z,\bar{z})$ and $(u,\bar{u})$, respectively; they are related by the transformation $z = u^{-1}$. 
In $N$, we can locally express $\omega$ in terms of a rational function $\varphi(z)$, known as the `twist function'\footnote{Note, that a M{\"o}bius transformation can always be performed such that there is a pole of $\omega$ located at $\infty$.}:
\begin{equation}
    \omega = \varphi(z) dz = \sum_{q \in P_0} \frac{k_q dz}{z-q} \ca
\end{equation}
where $P_0$ denotes the set of poles located in $N$.
We shall call the coefficients, $k_q$, of $\omega$ `levels' because of their appearance in a series of Kac-Moody current algebras in section \ref{subsec:gauge-fixing-step-one}. 
These levels can be determined in terms of the twist function by $k_q = \text{Res}_q \, \varphi(z)$, when $q \in P_0$, and $k_\infty = - \text{Res}_0 \, u^{-2} \varphi(u^{-1})$ should $\omega$ have a pole at $z=\infty$. 
Notice that the sum of all levels vanishes, $\sum_{q \in P} k_q = 0$, because the Riemman sphere is a compact manifold. 

The principal bundle $\mathcal{P}$ has two structures: its base $M$, which is the spacetime of our theory, and its fibre.  
The base is a four-dimensional manifold given by the cross-product of two surfaces, $M = W \times \mathbb{C} P^1_\omega$. 
The first of these, $W$, will eventually form the worldsheet of a class of $\sigma$-models discussed in \cite{Stedman:2021wrw}. 
For simplicity, we assume it is a cylinder of radius $R$, $W = S^1_R \times \mathbb{R}$, parametrised by the coordinates $x \in [- \pi R, \pi R)$ and $t \in \mathbb{R}$. 
The second surface $\mathbb{C} P^1_\omega$ is a punctured Riemann sphere given by $\mathbb C P^1_\omega = \mathbb CP^1 \setminus P \cup Z$. 
Since they will play a significant role in the following, we denote the copies of $W$ removed from $M$ at the punctures by $W_p := W_p \times (p , \bar p)$ for $p \in P \cup Z$. 
We call these `surface defects' and denote their collective set by $D = \sqcup_{p \in P \cup Z} W_p$. 

Throughout the following, we will encounter integrals with a `boundary term' of the form: 
\begin{equation}
    I = \frac{i}{8 \pi^2}\int_M d(\omega \wedge \eta) \ca
\end{equation} 
where $\eta \in \Omega^2 (M)$ is assumed to satisfy the following properties: 
\begin{enumerate} 
    \item For each $q \in P$ there exists a disc $D_q$ in which $\eta$ is continuous. If $q \in P_N$ then $D_q$ is the region defined by $ |z-q| \leq R_q$, while for the pole at $\infty$, it is instead $|z|^{-1} \leq R_\infty$; 
    \item For each $\zeta \in Z$ there exists a two-form $\eta^{(\zeta)} = (z-\zeta)^m \eta$, for $m=0,1$ or $2$, which is continuous in the disc $D_\zeta$ defined by $|z - \zeta| \leq R_\zeta$.
\end{enumerate} 
To calculate $I$, we expand each puncture $p \in P \cup Z$ to a small disc $B_p$ of radius $r_\zeta$, evaluate the resulting integrals over $\partial B_p \times W_p$ using a Fourier expansion of $\eta$, and finally, shrink the disc back to a point; see appendix \ref{appendix:Boundary-Integrals} for further details. 
At the end of this process, we find that: 
\begin{equation}
    I = \sum_{q \in P} \frac{k_q}{4 \pi} \int_{W_q} \iota^*_q \eta + \sum_{\zeta \in Z} \frac{k_\zeta}{4 \pi} \int_{W_q} \iota^*_\zeta \eta^{(\zeta)} \ca \label{eq:boundary-integral-result}
\end{equation}
where $k_\zeta := \varphi'(\zeta)$ and in which $\iota^*_p \eta$ is an abuse of notation that denotes the pullback of the boundary inclusion $\iota_p : \partial B_p \times W_p \rightarrow M$ in the zero radius limit. 
In appendix \ref{appendix:Boundary-Integrals}, we find that:
\begin{equation}
    \iota^*_q \eta = \eta(q) \ca \qquad \iota^*_\zeta\eta^{(\zeta)} = \res{\zeta} (z-\zeta) \, \eta \ca
\end{equation}
for $q \in P$ and $\zeta \in Z$ ($\iota_\zeta^*\eta = 0$ unless $m=2$). 

The structure of $\mathcal P$'s fibre is a little more subtle.
To construct it, we introduce two complex semi-simple Lie groups, $\mathcal{G}$ and $\mathcal{H} \subseteq \mathcal{G}$, and consider the string of inclusions $\mathcal{G \times H \supseteq H \times H \supset H_{\text{diag}}}$, in which $\mathcal{H}_{\text{diag}}$ is a diagonal copy of $\mathcal{H}$ within $\mathcal{H \times H}$. 
From this, we know that $\mathcal{G \times H}$ also contains a diagonal copy of the `common centre', \textit{i.e} the largest abelian subgroup of $\mathcal{H}$ whose elements commute with all of $\mathcal{G}$. 
We denote this by $\mathcal{Z}$. 
Written in terms of these three groups, the principal bundle $\mathcal{P}$ has the fibre: 
\begin{equation}
    \mathcal{G \times H / Z}
\end{equation}
where the modding out of $\mathcal{Z}$ follows from boundary conditions imposed at the points $P$. 
We will explain this in detail in the subsection \ref{section:common-centre}. 

Before we give the doubled action, it is necessary to recall a few important facts about the Lie algebras of $\mathcal{G}$ and $\mathcal{H}$, which we denote by $\lie{g}$ and $\lie{h}$, respectively. 
These are: 
\begin{itemize}
    \item There exists a (not necessarily unique) embedding of $\lie{h}$ into $\lie{g}$, $E : \lie{h} \rightarrow \lie{g}$. 
    Being explicit, given a pair of appropriately chosen bases $\{t_i \} \subset \lie h$ and $\{T_I \} \subset \lie g$ 
    we define $E$ such that $E(t_i) = T_i$, where the set $\{T_i\}$ forms a basis of $\lie{h} \hookrightarrow\lie{g}$. 
    As long as its action is restricted to $\lie h \hookrightarrow \lie g$, the inverse of $E$ is given by $E^{-1} (T_i) = t_i$. 
    
    \item  For each irreducible representation $R$ of $\lie{g}$ we can define a non-degenerate symmetric invariant bilinear form $\killing{\cdot}{\cdot}_R : \lie{g} \times \lie{g} \rightarrow \mathbb{C}$ which is proportional to $\lie{g}$'s Killing form:
    \begin{equation}
        \killing{X}{Y} := I_R K(X,Y) \ca \label{eq: Bilinear form def}
    \end{equation}
    where $I_R$ is the `index of the representation'. 
    The embedding map $E$ induces on $\lie{h}$ a representation $r = R \circ E$, implying the existence of a (non-degenerate symmetric) bilinear $\killing{\cdot}{\cdot}_{\lie h} : \lie{h} \times \lie{h} \rightarrow \mathbb{C}$ on $\lie{h}$ defined by: 
    \begin{equation}
        \killing{X}{Y}_{\lie h} := \frac{1}{I_{\lie h \hookrightarrow \lie g}}\killing{E(X)}{E(Y)} \ca \label{eq: h bilinear form}
    \end{equation}
    where $I_{\lie{h} \hookrightarrow \lie{g}}$ is the `index of embedding' \cite{Dynkin:semi-simple}. 
    The quotient by the common centre mentioned above imposes an additional requirement on the allowed pairs of representations of $\lie{g} \times \lie{h}$, as $(R,r)$ must be a representation of both $\mathcal{ G \times H}$ and $\mathcal{ G \times H / Z}$. 
    Therefore, we must choose $(R,r)$ such that the elements of $\lie{g} \times \lie{h}$ transform trivially under the action of $\mathcal{Z}$. 

    \item We can always perform a vector space decomposition $\lie g = \lie f \oplus \lie h$ in which the subspaces are orthogonal to each other $\killing{\lie f}{\lie h}=0$. 
    This guarantees the existence of a projection of $\lie g$ onto both $\lie h$ and $\lie f$, each respectively defined by: 
    \begin{equation}
    X^{\lie h} := \sum_{i} \killing{X}{T_i} T_i \ca \qquad X^\lie{f} := X - X^{\lie h} \ca 
    \end{equation}
    where the sum is over the basis of $\lie h \hookrightarrow \lie g$. 

    \item For both $\lie g$ and $\lie h$ there exists a pair of dual bases $\{T^I\}$ and $\{t^i\}$ which are defined to satisfy $\killing{T_I}{T^J} = \delta_I^{\ J}$ and $\killing{t_i}{t^j}_{\lie h} = \delta_i^{\ j}$. 

\end{itemize}
We will restrict ourselves to considering pairs of groups $(\mathcal{G},\mathcal{H})$ for which the coset $\mathcal{G/H}$ is a reductive homogeneous space \cite{Kobayashi1963}. 
When expressed as a condition on the subspaces $\lie f$ and $\lie h$, this is the requirement that $\lie f$ be invariant under the adjoint action of $H$, \textit{i.e} $v \lie f v^{-1} \subset \lie f$ for all $v \in H$. 

As it will be useful throughout the following, when $\eta = \killing{X}{Y}$ in equation \cref{eq:boundary-integral-result} we will use the condensed notation: 
\begin{equation}
    \tr{X}{Y}{q} := k_q \, \killing{\iota_q^* X}{\iota_q^* Y} \ca \quad \tr{X}{Y}{\zeta} := k_\zeta \, \iota_\zeta^* \langle X , Y \rangle^{(\zeta)} \ca \label{eq:normalised-bilinear}
\end{equation}
for $q \in P$ and $\zeta \in Z$, and under the assumption that $\killing{X}{Y}$ has at most a second degree pole at $\zeta \in Z$. 

The doubled 4dCS action is constructed as follows. 
On the bundle $\mathcal{P}$, there exists a connection that can be decomposed into the pair $(A,B) \in \lie{ g \times h}$ in the representation $(R,r)$ subject to the conditions discussed above. 
Under a gauge transformation generated by $(u,v) \in \mathcal{ G \times H / Z}$ these gauge fields transform in the usual way: 
\begin{equation}
    A \rightarrow {}^u A = u (d + A) u^{-1} \ca \qquad B \rightarrow {}^v B = v (d + B) v^{-1} \fp
\end{equation} 
Using the fields $A$, $B$ and one-form $\omega$, we construct the doubled 4dCS action: 
\begin{equation}
    S' (A,B) = \frac{i}{8 \pi^2} \int_{M} \! \! \! \omega \wedge \left( \text{CS}_\lie{g}(A) - I_{\lie{h \hookrightarrow g}} \text{CS}_\lie{h}(B) \right) + \frac{I_{\lie{h \hookrightarrow g}}}{4 \pi} \sum_{q \in P} \int_{W_q} \! \! \tr{E^{-1} (A^{\lie h} )}{B}{q}{}_{\lie{h}}
\end{equation}
where $\text{CS}_\lie{g}(A)$ and $\text{CS}_\lie{h}(B)$ are the Chern-Simons three-forms:
\begin{equation}
    \text{CS}_\lie{g}(A) = \killing{A}{dA + \tfrac{1}{3} [A,A]} \ca \qquad \text{CS}_\lie{h}(B) = \killing{B}{dB + \tfrac{1}{3} [B,B]}_\lie{h} \ca 
\end{equation}
in which $[A,A] = [A_\mu,A_\nu] d x^\mu \wedge d x^\nu$. 
The above action is invariant under the transformations $A \rightarrow A + \chi dz$ and $B \rightarrow B + \xi dz$ due to the wedge product with $\omega$. 
We use this symmetry to set $A_{z} = B_{z} = 0$. 

Using the embedding $E$, and equation \cref{eq: h bilinear form} $S'$ is: 
\begin{align}
    S'(A,B) &= \frac{i}{8 \pi^2} \int_{M} \! \! \! \omega \wedge \left( \text{CS}_{\lie g}(A) - \text{CS}_\lie{g}(B) \right) + \frac{1}{4 \pi} \sum_{q \in P} \int_{W_q} \! \! \! \! \tr{A}{B}{q} \ca \label{eq: unimproved action}
\end{align} 
where we have abused notation by also denoting the embedding of $B$ into $\lie{g}$ by $B$. 
From here on in, we regard $B$ as being $\lie{g}$-valued with vanishing $\lie{f}$ components.

\subsection{Functional Differentiability and Boundary Terms \label{subsec: Functional differentiability}} 

When performing the Hamiltonian or Lagrangian analysis of a field theory on a manifold with boundary, one will encounter functionals $F[f]$, whose variations $\delta F[f]$ contain non-trivial boundary terms $\delta F_{\partial}[f]$. 
The presence of these terms means the functional derivative of $F[f]$ is ill-defined, making a consistent analysis of the theory impossible.
Thankfully, this issue can be solved by following Regge and Teitelboim's prescription developed in \cite{Regge:1974zd} and reviewed in \cite{Banados:2016zim} - see also \cite{Banados:1994tn,Banados:1998gg} for a discussion of its relevance in three-dimensional Chern-Simons.
One starts by choosing a set of boundary conditions that constrain the theory's fields, but importantly, not their variations. 
These are then imposed upon $\delta F[f]$, hopefully removing several boundary terms in the process. 
Should there be any remaining terms, one must either impose further boundary conditions or `improve' the functional. 
To do this one adds new boundary terms to $F[f]$, producing $F_I[f] = F[f] + B[f]$, such that the variation of $B[f]$ cancels with the remaining terms of $\delta F_{\partial}[f]$ after imposing the boundary condition.
Physically speaking, this guarantees that the improved functional $F_I[f]$ is associated with a single space of field configurations $\mathscr{Q}$, and thus phase space $\mathscr{P} = T^* \mathscr{Q}$. 
If a functional's variation contains no boundary terms upon imposing any boundary conditions, we say it is `functionally differentiable'. 

\begin{table}[h!]
    \centering
    \def\arraystretch{1.5}
    \begin{tabular}{c|c}
        Defect Type & Boundary Conditions \\
        \hline 
        Gauged Chiral & $A_{t,x}^{\lie{h}}(q,\bar{q}) = B_{t,x}(q,\bar{q})$, $A_{t}^{\lie{f}}(q,\bar{q}) = A_{x}^{\lie{f}}(q,\bar{q})$ \\
        Gauged Anti-chiral & $\ \ A_{t,x}^{\lie{h}}(q,\bar{q}) = B_{t,x}(q,\bar{q})$, $A_{t}^{\lie{f}}(q,\bar{q}) = -A_{x}^{\lie{f}}(q,\bar{q})$
    \end{tabular}
    \caption{The gauged chiral and anti-chiral boundary conditions of \cite{Stedman:2021wrw}}
    \label{tab:boundary-conditions}
\end{table} 

To illustrate this concept, consider the variation of the action $S'(A,B)$ subject to the boundary conditions given in Table \ref{tab:boundary-conditions} and under the assumption that both $A_W = A_x dx + A_t dt$ and $B_W = B_x dx + B_t dt $ may have simple poles at the points in $Z$. 
In this example, one encounters a boundary term to which we apply \cref{eq:boundary-integral-result} and find: 
\begin{equation}
    \delta S'_{\partial M} = \sum_{q \in P} \delta S'_q + \sum_{\zeta \in Z} \delta S'_{\zeta} \ca
\end{equation}
where:
\begin{equation}
    \delta S'_q = \frac{1}{4 \pi} \int_{W_q} \! \! \tr{A-B}{\delta A+ \delta B}{q} \ca \quad  \delta S'_{\zeta} = \frac{1}{4\pi} \int_{W_\zeta} \! \! \left( \tr{A}{\delta A}{\zeta} - \tr{B}{\delta B}{\zeta} \right) \fp
\end{equation} 

If we assume that a gauged chiral boundary condition is imposed at $q \in P$ and apply the Regge-Teitelboim prescription to $\delta S'_q$ we find\footnote{Notice that boundary term in \cref{eq: unimproved action} already gives us an improvement term since it enables us to impose the boundary condition $A^\lie{h}_{t,x} = B_{t,x} \fp$}:
\begin{equation}
    \delta S'_{q} = - \frac{1}{4\pi} \int_{W_q}\! \! \tr{A_x^\lie{f}}{\delta A_t^\lie{f} - \delta A_x^\lie{f}}{q} \, d^2x \ca
\end{equation}
where $d^2x = dt \wedge dx \,$.
Following their approach, we must either impose the additional boundary conditions, $A_{x}^{\lie{f}}(q,\bar{q}) = 0$, or, if we wish to continue working with the gauged chiral condition, improve the action. 
We opt for the latter approach as the additional conditions required of the former are stronger than we would like. 
To do this, we add the improvement term:
\begin{equation}
    I_q^{+} = \frac{1}{4 \pi} \int_{W_q}\! \! \tr{A_x^\lie{f}}{A_{t}^{\lie{f}}-A_x^{\lie{f}}}{q}  \, d^2 x \ca
\end{equation} 
whose variation cancels with $\delta S'_{q}$ once the boundary conditions are imposed. 
The same analysis can be done with the gauged anti-chiral conditions substituted for the chiral ones, in which case one finds the improvement term: 
\begin{equation}
    I_q^{-} = \frac{1}{4\pi} \int_{W_q} \! \! \tr{A_x^\lie{f}}{A_t^{\lie{f}}+A_x^\lie{f}}{q} \, d^2x \fp
\end{equation} 

To perform the same analysis for the $\delta S'_{\zeta}$ we divide the set of zeros $Z$ into two pairs of disjoint subsets, $Z_{\pm}^{A}$, and, $Z_{\pm}^B$ such that $Z_{+}^A \sqcup Z_{-}^A = Z_{+}^B \sqcup Z_{-}^B = Z$. 
Using this splitting we define for each $\zeta$ a pair of numbers $(\alpha_{\zeta}, \beta_{\zeta})$ via the conditions $\alpha_{\zeta} = \pm 1$ if $\zeta \in Z_{\pm}^A$ and $\beta_{\zeta} = \pm 1$ if $\zeta \in Z_{\pm}^B$. 
The reader should think of this as the defining data for a class of defects located at the zeros. 
Using $(\alpha_{\zeta}, \beta_{\zeta})$ we construct the improvement term:
\begin{equation}
    I_{\zeta} = \frac{1}{4 \pi} \int_{W_\zeta} \! \! \tr{A_x}{A_t - \alpha_\zeta A_x}{\zeta} \, d^2x  - \frac{1}{4 \pi} \int_{W_\zeta}  \! \! \tr{B_x}{B_t - \beta_\zeta B_x}{\zeta} \, d^2x \fp
\end{equation} 
Once these are added to the action, the boundary term in the variation is modified to: 
\begin{equation}
    \delta S'_{\zeta} + \delta I_\zeta = \frac{1}{4 \pi} \int_{W_\zeta} \! \! \left( \tr{A_t - \alpha_\zeta A_x}{\delta A_x}{\zeta} + \tr{B_t - \beta_\zeta B_x}{\delta B_x}{\zeta} \, \right) d^2 x \fp \label{eq: variation of the action from a zero}
\end{equation}
The requirement that this term vanishes fixes a particular pole structure of $A_t-\alpha_{\zeta} A_x$ and $B_t-\beta_{\zeta} B_x$ in $z$. 
To see this we suppose that the worldsheet components of the two gauge fields behave near $\zeta$ as $A_{x,t}(z,\bar{z}) = \tilde{A}_{x,t}(z,\bar{z})/(z-\zeta)$ and $B_{x,t}(z,\bar{z}) = \tilde{B}_{x,t}(z,\bar{z})/(z-\zeta)$ with $\tilde{A}_{x,t}(z,\bar{z})$ and $\tilde{B}_{x,t}(z,\bar{z})$ regular at $z = \zeta$. 
If we substitute these two expressions into equation \cref{eq: variation of the action from a zero}, evaluate the residue and require the result vanishes one finds that $\tilde{A_{t}}(\zeta,\bar{\zeta}) - \alpha_{\zeta} \tilde A_{x}(\zeta,\bar{\zeta}) = \tilde{B_{t}}(\zeta,\bar{\zeta}) - \beta_{\zeta} \tilde B_{x}(\zeta,\bar{\zeta}) = 0$. 
Thus, we have the boundary conditions: 
\begin{itemize}
    \item Chiral pole: when $\zeta \in Z_{+}^A,Z_{+}^B$ we require that $A_t(\zeta,\bar{\zeta}) - A_x(\zeta,\bar{\zeta})$ and $B_t(\zeta,\bar{\zeta}) - B_x(\zeta,\bar{\zeta})$ are regular, 
    \item Anti-chiral pole: while if $\zeta \in Z_{-}^B,Z_{-}^B$ we instead ask that $A_t(\zeta,\bar{\zeta}) + A_x(\zeta,\bar{\zeta})$ and $B_t(\zeta,\bar{\zeta}) + B_x(\zeta,\bar{\zeta})$ are regular. 
\end{itemize} 

Hence, after following Regge and Teitelboim's prescription, we find the improved action: 
\begin{equation}
    S(A,B) = S'(A,B) + \sum_{q \in P_{+}} I_q^+ + \sum_{q \in P_{-}} I_q^- + \sum_{\zeta \in Z} I_\zeta \ca
\end{equation}
where $P_{+} \subseteq P$ (resp. $P_{-}$) is the subset of the poles at which the gauged (anti)-chiral boundary conditions are imposed. 
Note that when fixing the boundary terms, we have implicitly chosen a metric on $W_q$. 
It is this choice that determines whether the $\sigma$-models produced by the construction are Euclidean or Lorentzian. 

\subsection{The Common Centre $\mathcal{Z}$\label{section:common-centre}}

When defining the bundle $\mathcal{P}$ we claimed the gauge group of our theory is $\mathcal{ G \times H / Z}$, but did not explain the origins of the quotient. 
It is this that we now turn to.
We do this by following an argument similar to one found in \cite{Chu:1991pn}, although in a different context. 

The boundary condition $A_{t,x}^{\lie{h}}(q,\bar{q}) = B_{t,x}(q,\bar{q})$ must be preserved under a gauge transformation generated by $(u,v)$. 
This implies the following constraint on the generators at $q \in P$:
\begin{equation}
    v^{-1}_q u_q \in C(\lie{h}) \times N(\lie{f}) \ca \label{eq: bc constraint}
\end{equation}
where $v_q = v(q,\bar{q})$ and $u_q = u(q,\bar{q})$. 
The groups $C(\lie{h})$ and $N(\lie{f})$ in the above expression are, respectively, the centraliser and normaliser of $\lie{h}$ and $\lie{f}$ within $\mathcal G$ -- the derivation of this constraint can be found in \cite{Stedman:2021wrw}. 
The above equation is invariant under two transformations. 
The first of these is a monodromy transformation $(M,M) \in \mathcal{H}_{\text{diag}}$ that maps $u_q(x + 2 \pi R)$ and $v(x+2\pi R)$ to $u_q$ and $v_q$ as we wind around the cylinder $W_q$. 
It is given by: 
\begin{equation}
    u_q(x + 2 \pi R) = M u_q(x) \ca \qquad v_q (x+2 \pi R) = M v_q (x) \ca
\end{equation}
where $M$ is constant. 
The second transformation arises because equation \cref{eq: bc constraint} is invariant under the shift: 
\begin{equation}
    (u_q,v_q) \mapsto (h_q,h_q) \cdot (u_q,v_q) = (h_q u_q,h_q v_q) \ca \label{eq: ambiguity in gauge trans}
\end{equation}
where $(h_q,h_q) \in \mathcal{H}_{\text{diag}}$ is again constant. 
This shift provides us with the map $M \mapsto h_q^{-1} M h_q$ which can be used to rotate $M$ into a maximal torus $\mathcal{T_H}$ of $\mathcal{H}$, fixing it up to a Weyl group transformation. 

After fixing the monodromy of $(u_q,v_q)$ in this way, the group of symmetries in equation \cref{eq: ambiguity in gauge trans} is reduced so that $(h_q,h_q)$ is an element of a maximal torus $\mathcal{T_{H_{\text{diag}}}}$ of $\mathcal{H_{\text{diag}}}$.
Now, since the common centre is the largest subgroup of $\mathcal{T_{H_{\text{diag}}}}$ that is also a normal subgroup of $\mathcal{G \times H}$ it follows that $\mathcal{G \times H} \cong \mathcal{Z \rtimes \left( G \times H/Z \right)}$. 
Using this fact we can express the generators of the gauge transformations by $(u_q,v_q) = (a,a) \cdot (\tilde{u}_q,\tilde{v}_q)$, where $(a,a) \in \mathcal{Z}$ and $(\tilde{u}_q,\tilde{v}_q) \in \mathcal{G \times H/Z}$, and use the reduced symmetry given by \cref{eq: ambiguity in gauge trans} to remove the $(a,a)$ factor. 
This implies our gauge symmetry group is $\mathcal{G \times H/Z}$. 

\section{The Bulk Hamiltonian Analysis \label{sec: Bulk Hamiltonian analysis}}

The overarching goal of this section is to perform the Hamiltonian analysis of doubled 4dCS; it is split into four parts. 
In the first subsection, we derive the theory's Hamiltonian, primary constraints and Poisson brackets. 
In the second part, we follow Dirac's algorithm for deriving the first-class constraints of a gauge theory, and in the third part, we impose a series of gauge choices. 
We will conclude by deriving the Poisson brackets on the associated constraint surface. 

Throughout this and succeeding sections, we use a tensor notation to simplify tensor products of objects valued in $\lie{g}$. 
In particular, a subscripted index of the form $\textbf{i}$ is used to indicate that a matrix is contracted with a generator of $\lie{g}$ in the $i$-th factor, for example $A_{\textbf{1}} = A^{I} T_I \otimes I$, where $I$ is the identity. 
When the matrix is multi-dimensional, the ordering of these indices indicates which component the $i$-th factor is contracted with, \textit{i.e.} $M_{\textbf{12}} = M^{ij} T_i \otimes T_j$ or $M_{\textbf{21}} = M^{ji} T_i \otimes T_j$. 
Importantly, naively different orderings will often be related to each other due to the symmetry properties of the contracting matrix, for instance $M_{\textbf{12}} = M_{\textbf{21}}$ if $M^{ij}$ is symmetric, or $M_{\textbf{12}} = - M_{\textbf{21}}$ if it is anti-symmetric. 
When the matrix is totally symmetric, we use a sans-serif font and will occasionally drop the superfluous indices \textit{e.g.} $\mathsf{M} = M_{\textbf{12}} = M_{\textbf{21}}$.

\subsection{The Hamiltonian} 

We pass from the Lagrangian formalism to the Hamiltonian one in three steps.
In the first, we take $t \in \mathbb R$ to be our time coordinate and foliate $M$ with a set of spacelike hypersurfaces, each given by $\Sigma := S^1_R \times \mathbb CP^1$ so that $M = \mathbb R \times \Sigma$.  
This enables us to decompose the exterior derivative and fields into: 
\begin{equation}
    A = A_t dt + A_\Sigma \ca \quad B = B_t dt + B_\Sigma \ca \qquad d= dt \partial_t + d_{\Sigma} \fp 
\end{equation} 
By substituting these expressions into the Chern-Simons three-form, we find:
\begin{equation}
    \omega \wedge\text{CS}(A) = dt \wedge \omega \wedge \langle A_{\Sigma}, \partial_t A_{\Sigma} \rangle - dt \wedge \omega \wedge \langle A_t, 2 F_{\Sigma}(A) \rangle - d_{\Sigma} \left( \omega \wedge \langle A_t dt, A_{\Sigma}\rangle \right) \ca \label{eq:CS-3form-rewritten}
\end{equation}
after using the product rule, $d_{\Sigma} \omega = 0$, and in which $F_{\Sigma}(A) = d_{\Sigma} A_{\Sigma} + \frac{1}{2} [A_{\Sigma},A_{\Sigma}]$. 
An analogous expression for $\omega \wedge \text{CS}(B)$ can be found by substituting $A$ for $B$ in \cref{eq:CS-3form-rewritten}. 

In the second step, we plug \cref{eq:CS-3form-rewritten}, and its equivalent for $\omega \wedge \text{CS}(B)$ into the improved action $S(A,B)$ and apply 
\cref{eq:boundary-integral-result} to simplify the boundary term. 
This yields $S(A,B) = S_M -\sum_{q \in P} S_q - \sum_{\zeta \in Z} S_{\zeta}$, where: 
\begin{equation}
    S_{M} = \frac{i}{4 \pi^2} \int_{M} \! \! \!dt \wedge \omega \wedge \left( \frac{1}{2} \langle A_{\Sigma}, \partial_t A_{\Sigma}\rangle - \frac{1}{2} \langle B_{\Sigma}, \partial_t B_{\Sigma} \rangle -\langle A_t, F_{\Sigma}(A) \rangle + \langle B_t, F_{\Sigma}(B)\rangle \right) \ca
\end{equation}
and:  
\begin{align}
    S_q &= \frac{1}{4 \pi} \int_{W_q} \! \! \left( \tr{A_t^\lie{h} +B_t}{A_x^{\lie{h}} - B_x}{q} + \epsilon_q \tr{A_x^\lie{f}}{A_x^\lie{f}}{q} \right) \, d^2 x \ca \\
    S_\zeta &= \frac{1}{4 \pi} \int_{W_\zeta} \! \! \left( \alpha_\zeta \tr{A_x}{A_x}{\zeta} - \beta_\zeta \tr{B_x}{B_x}{\zeta} \right) \, d^2 x\ca
\end{align}
in which $\epsilon_q = \pm 1$ for $q \in P_{\pm}$. 

In the final step, we find the Hamiltonian by performing a Legendre transformation of the Lagrangian. 
This requires the use of the conjugate momenta for $A_{\mu}$ and $B_{\mu}$ which are denoted by $P^{\mu}$ and $Q^{\mu}$, respectively. 
While there are several methods for finding these momenta, we choose to read them off of the theory's symplectic potential $\mathscr{A}$ as it will reappear when we calculate Poisson brackets shortly. 
The potential is easily found from an action's variation where it appears as the differentiand of the total derivative of $t$, \textit{i.e.} $\int_M \partial_t \mathscr A$. 
For the doubled theory, it is: 
\begin{equation}
    \mathscr{A} := \int_\Sigma \! \left(  \killing{P^{\mu}}{\delta A_{\mu}} + \killing{Q^{\mu}}{\delta B_{\mu}} \right) \, d^3x  = \frac{i}{8 \pi^2} \int_\Sigma \! \omega \wedge ( \langle A_{\Sigma}, \delta A_{\Sigma} \rangle - \langle B_{\Sigma}, \delta B_{\Sigma} \rangle ) \ca \label{eq: symplectic potential}
\end{equation} 
where $d^3x = dx \wedge d \bar{z} \wedge dz$. 
Hence, the conjugate momenta are:
\begin{equation}
    P^a = - \frac{i \varphi}{8 \pi^2} \epsilon^{ab} A_b\ca \quad P^t = 0 \ca \quad Q^a = \frac{i \varphi}{8 \pi^2} \epsilon^{ab} B_b \ca \quad Q^t = 0 \ca \label{eq: conjugate momenta}
\end{equation}
where $a,b=x,\bar z$ and $\epsilon^{x \bar{z}} = - \epsilon^{\bar{z} x} = 1$. 
A simple calculation leads us to the Hamiltonian: 
\begin{align}
    H = H_{\Sigma}[A_t,B_t] + \sum_{q \in P} ( H_q^{\lie{h}}[A_t,B_t] + H_{q}^{\lie{f}}[A_x] ) + \sum_{\zeta \in Z} H_{\zeta}[A_x , B_x]
\end{align}
where:
\begin{align} 
    H_{\Sigma}[A_t,B_t] &= \frac{i}{4 \pi^2} \int_\Sigma \! \omega \wedge ( \langle A_t, F_{\Sigma}(A) \rangle  - \langle B_t, F_{\Sigma}(B) \rangle ) \ca \\
    H_{q}^{\lie{h}}[A_t,B_t] &= \frac{1}{4 \pi} \int  \tr{A_t^\lie{h} +B_t}{A_x^{\lie{h}} - B_x}{q} \, dx \ca \\
    H_{q}^{\lie{f}}[A_x] &= \frac{\epsilon_q}{4 \pi} \int \tr{A_x^\lie{f}}{A_x^\lie{f}}{q} dx \ca \\ 
    H_{\zeta}[A_x,B_x] &= \frac{1}{4 \pi} \int \left( \alpha_\zeta \tr{A_x}{A_x}{\zeta} - \beta_\zeta \tr{B_x}{B_x}{\zeta} \right) \, dx \fp
\end{align} 

\subsection{Poisson Brackets and Symplectic Two-Form Inversion\label{section:symplectic-inversion}} 

As similar calculations will appear throughout the following, let us discuss how a set of Poisson brackets is found from a symplectic two-form: 
\begin{equation}
    \Omega = \frac{1}{2} \int_{\Sigma} \int_{\Sigma} \Omega_{ab}(x,y) \killing{\delta X^a(x)}{\delta X^b(y)} \, d^3 x \, d^3 y \ca
\end{equation} 
in which $X^a$ for $a=1,\ldots, \dim_{\mathbb R} \mathscr P$ is a $\lie g$-valued local coordinate on a phase space $\mathscr{P}$. 
We will call this process `inversion' of $\Omega$. 
Note that we will often use $x$ and $y$ to indicate $(x,z,\bar z)$ and $(y,w,\bar w)$, as we have done in the above equation, so that our notation does not become too cumbersome. 

The starting point of inversion is to replace $\killing{\delta X^a(x)}{\delta X^b(y)}$ in the integrand of $\Omega$ with the split Casimir $\mathsf{C} = T_I \otimes T^I = T^I \otimes T_I$ where the sum over $I$ is implied. 
This produces an $\lie g \otimes \lie g$ valued matrix with the components $\Omega_{ab}(x,y) \, \mathsf{C}$ that, by an abuse of notation, is also denoted by $\Omega$. 
The inverse $\Omega^{-1}$ of this matrix is defined to satisfy the condition: 
\begin{equation}
    \int_{\Sigma} \killing{\Omega_{\textbf{12}}(x,y')}{\Omega_{\textbf{23}}^{-1}(y',y)} \, d^3 y' = \mathsf{C}_{\textbf{13}}\delta^3(x-y) \ca
\end{equation}
and the Poisson brackets of $X^a$ are given by: 
\begin{equation}
    \{ X^a (x) , X^b (y) \} := - \Omega^{-1}_{ab}(x,y) \ca
\end{equation} 
where the minus sign appears as a matter of convention. 

For the doubled theory, the initial symplectic two-form is: 
\begin{align}
    \Omega :=& \int_{\Sigma} \left( \killing{\delta A_{\mu}}{\delta P^\mu} + \killing{\delta B_{\mu}}{\delta Q^\mu} \right) d^3 x  \\
    =& \frac{1}{2} \int_{\Sigma} \int_{\Sigma} \delta(x-y) \left(  \killing{\delta A_{\mu}(x)}{\delta P^\mu(y)} - \killing{\delta P^\mu(x)}{\delta A_{\mu}(y)} \right. \\
    & \hspace{1.5cm}  + \left. \killing{\delta B_{\mu}(x)}{\delta Q^\mu(y)}  - \killing{\delta Q^\mu(x)}{\delta B_{\mu}(y)} \right) d^3 x \, d^3 y \ca \nonumber
\end{align}
which, via inversion, produces the usual Poisson brackets of the canonical variables: 
\begin{equation}
    \{ A_{\mu \textbf{1}}(x) , P^\nu_{\textbf{2}}(y) \} = \delta_{\mu}^{\ \, \nu} \, \mathsf{C} \,  \delta^3 (x-y) \ca \qquad \{ B_{\mu \textbf{1}}(x) , Q^\nu_{\textbf{2}}(y) \} = \delta_{\mu}^{\ \, \nu} \, \mathsf{C}^{\lie{h}} \, \delta^3 (x-y) \ca \label{eq:initial-brackets}
\end{equation} 
where $\mathsf C^\lie h = T_i \otimes T^i$ is the projection of $\mathsf C$ into $\lie h$. 

\subsection{Dirac's Algorithm and the Primary Constraints} 

Our aim for the rest of this section is to use Dirac's algorithm \cite{Dirac-1964} to perform the Hamiltonian analysis of the Doubled theory. 
This starts with the observation that gauge symmetry implies a redundancy in the canonical variables, as (some of) the conjugate momenta $P^I$ are determined by expressions involving the fields $\Phi_I$ and not their time derivatives, just as \cref{eq: conjugate momenta}. 
The goal of the algorithm is to eliminate any redundant variables. 

To do this, we introduce a set $\mathscr{C}$ of `constraints'. 
These are equations of the form $C_i (P^I, \Phi_I) = 0$ that we want the canonical variables to satisfy, and to each of which there is an associated submanifold $\mathscr{P}_i$ of the phase space called the `constraint surface'. 
When expressed in terms of these surfaces, the process of eliminating variables corresponds to the reduction of the phase space down to $\mathscr{P}(\mathscr{C}) := \mathscr P_1 \cap \cdots \cap \mathscr P_{|\mathscr C|}$. 
More often than not, we must proceed in several steps by introducing a string of inclusions for the sets of constraints and their associated surfaces: 
\begin{equation}
    \mathscr{C}_m \subset \cdots \subset \mathscr{C}_1 \subset \mathscr{C} \ca \qquad \mathscr{P}(\mathscr{C}) \subset \cdots \subset \mathscr{P}(\mathscr{C}_1) \subset \mathscr{P}(\mathscr{C}_m) \fp 
\end{equation} 
It is therefore often helpful to separate the constraints $\mathscr C$ into those that are satisfied by the remaining variables on the surface $\mathscr P (\mathscr C_i)$ and those that are not. 
Those in the former class are said to hold `strongly' and expressed as $C_i = 0$; otherwise, a constraint is said to hold `weakly' and $C_i \approx 0$ is used. 

There exist (potentially trivial) linear combinations of the constraints within $\mathscr C$ that are characterised by whether or not they weakly commute with all the constraints in $\mathscr C$, \textit{i.e.}\ if $\{C , C_i \} \approx 0$ for all $i$. 
When a constraint does satisfy this condition, it is said to be `first-class', otherwise, it is called `second-class'. 
The presence of first-class constraints has two consequences, the first of which is the existence of new constraints that further reduce the theory. 
These arise from the requirement that $C_{1st} \approx 0$ holds for all time, and are thus given by $\{C_{1st} ,H \} \approx 0$. 
Of course, the new constraint could also be first-class, in which case this step is repeated until no new constraints are found. 

The second consequence is that the symplectic form $\Omega_\mathscr C$ on $\mathscr{P}(\mathscr C)$ will be non-degenerate if $\mathscr C$ contains a first-class constraint. 
This makes it impossible to find a set of Poisson brackets for $\mathscr{P}(\mathscr C)$. 
Assuming that we have a complete set of constraints, this issue can be solved in two different ways. 
In the first, we make a series of gauge choices, which are auxiliary constraints that render a first-class constraint second-class. 
Once this is done, it becomes possible to impose all of the constraints strongly and find the Poisson brackets. 
The second approach requires that we work with gauge-invariant quantities, which are defined to commute with first-class constraints. 
The Poisson brackets between these objects will remain unchanged from their initial form. 
Both methods will feature in this work. 

Our starting point is to introduce a set of `primary constraints', collectively denoted by $\mathscr C_1$, that the momenta $(P^\mu,Q^\mu)$ and fields $(A_\mu,B_\mu)$ must satisfy. 
These are: 
\begin{equation}
    C^a = P^a + \frac{i \varphi}{8 \pi^2} \epsilon^{ab} A_b \approx 0 \ca \quad K^a = Q^a - \frac{i \varphi}{8 \pi^2} \epsilon^{ab}B_b \approx 0 \ca \quad P^t \approx 0 \ca \quad Q^t \approx 0 \ca
\end{equation} 
for $a,b = x, \bar{z}$. 
Next, we determine which of these are first or second-class by calculating their Poisson brackets with each other. 
Using \cref{eq:initial-brackets} we find, after integrating against a test function, that the only non-zero brackets between the constraints are: 
\begin{equation}
    \{ C^a_{\textbf{1}} (x),C^b_\textbf{2} (y)\} =  \frac{i \varphi(w)}{4 \pi^2} \epsilon^{ab} \, \mathsf{C} \, \delta^3 (x-y) \ca \quad \{ K^a_{\textbf{1}} (x),K^b_\textbf{2} (y)\} = -  \frac{i \varphi(w)}{4 \pi^2} \epsilon^{ab} \, \mathsf{C}^{\lie{h}} \, \delta^3 (x-y) \fp
\end{equation} 
This implies that $P^t \approx 0$ and $Q^t\approx 0$ are both first-class and that $C^a \approx 0$ and $K^b\approx 0$ are second. 

A set of second-class constraints $\mathscr C_1^{\text{2nd}}  \subset \mathscr C_1$ provides us with a series of relations amongst the canonical variables that hold weakly. 
By solving the constraints so that they hold strongly, we can eliminate some of the variables and reduce the phase space $\mathscr P$ down to the associated constraint surface $\mathscr P(\mathscr C_1^{\text{2nd}})$. 
When doing this, consistency with the now strong constraints requires that we use a new set of Poisson brackets on $\mathscr P(\mathscr C_1^{\text{2nd}})$, called `Dirac Brackets'. 
There are several different ways to find Dirac brackets, the most famous of which is a formula found in \cite{Dirac-1964} that projects the Poisson brackets of $\mathscr P$ down to a set on $\mathscr P(\mathscr C_1^{\text{2nd}})$. 
We will not use this formula. 
Instead, we modify the symplectic potential $\mathscr A$ to account for the fact that $P^t \approx 0$ and $Q^t \approx 0$ only hold weakly: 
\begin{equation}
    \mathscr{A}_1 = \mathscr A + \int_\Sigma \! \left( \killing{P^t}{\delta A_t} + \killing{Q^t}{\delta B_t} \right) \, d^3x \ca
\end{equation} 
and then find a symplectic two-form $\Omega^{(1)}$ on $\mathscr P(\mathscr C_1^{\text{2nd}})$ via the formula $\Omega^{(1)} := - \delta \mathscr A_1$ which yields: 
\begin{align}
    \Omega^{(1)} = \int_\Sigma \! \left( \frac{i \varphi}{4 \pi^2} \killing{\delta A_{\bar{z}}}{\delta A_x} - \frac{i \varphi}{4 \pi^2} \killing{\delta B_{\bar{z}}}{\delta B_{x}} + \killing{\delta A_t}{\delta P^t} + \killing{\delta B_t}{\delta Q^t} \right) \,  d^3 x \fp
\end{align}
The Dirac brackets for $\mathscr P(\mathscr C_1^{\text{2nd}})$ then follow from the inversion procedure described in section \ref{section:symplectic-inversion}: 
\begin{align}
    \{ A_{\bar{z} \textbf{1}}(x) , A_{x \textbf{2}}(y) \}_{(1)} &= - \frac{4 \pi^2 i}{\varphi(w)} \, \mathsf{C} \, \delta^3 (x-y) \ca & \{ B_{\bar{z} \textbf{1}}(x) , B_{x \textbf{2}}(y) \}_{(1)} &= \frac{4 \pi^2 i}{\varphi(w)} \, \mathsf{C}^{\lie{h}} \, \delta^3 (x-y) \label{eq: Dirac bracket 1}\\
    \{ A_{t \textbf{1}} (x) , P^t_{\textbf{2}}(y) \}_{(1)} &= \, \mathsf{C} \, \delta^3(x-y) \ca & \{ B_{t \textbf{1}} (x) , Q^t_{\textbf{2}}(y) \}_{(1)} &= \, \mathsf{C}^{\lie{h}} \, \delta^3(x-y) \fp
\end{align}
Later on, we will use an alternative approach, in which the constraints are imposed upon the symplectic form $\Omega^{(i)}$ of $\mathscr P (\mathscr C_{i})$ to produce a new two-form $\Omega^{(i+1)}$ on $\mathscr P (\mathscr C_{i+1})$ whose Poisson brackets will again be found via inversion. 

\subsection{Secondary Constraints and Proper Gauge Transformations} 

The analysis of the previous section leaves us with a set of first-class primary constraints $\mathscr C_1^{\text{1st}}=\{ P^t \approx 0, Q^t \approx 0 \}$. 
During our earlier description of Dirac's algorithm, we mentioned that first-class constraints must hold for all time, and that this requirement can introduce new `secondary' constraints. 
For the doubled theory, these new constraints are the Gauss laws: 
\begin{equation}
    \dot P^t = \{P^t(x),H(y)\}_1 = \frac{i \varphi(z)}{4 \pi^2} F(A) \approx 0 \ca \qquad \dot Q^t =\{Q^t(x),H(y)\}_1 = - \frac{i \varphi(z)}{4 \pi^2} G(B) \approx 0 \ca \label{eq: secondary constraints 1}
\end{equation} 
where\footnote{We use $'$ and $\bar \partial$ to indicate the derivatives with respect to $x$ and $\bar z$.} $F(A) = \bar{\partial} A_x - A_{\bar{z}}' + [A_{\bar{z}},A_{x}]$ and $G(B) = \bar{\partial} B_x - B_{\bar{z}}' + [B_{\bar{z}},B_{x}]$. 
We shall use $\mathscr{C}_2$ to denote the now expanded set of constraints:
\begin{equation}
    P^t \approx 0\ca \quad  Q^t \approx 0 \ca \quad \varphi(z) F \approx 0 \ca \quad \varphi(z) G \approx 0 \ca
\end{equation}
whose associated constraint surface is $\mathscr{P}(\mathscr{C}_2)$. 

In the previous subsection, we introduced the notions of first- and second-class constraints and separated the primary constraints accordingly. 
We now aim to do the same for the secondary constraints. 
For $P^t \approx 0$ and $Q^t \approx 0$ this is trivial, as none of the constraints contain $A_t$ or $B_t$, and so they remain first-class. 
For the Gauss law constraints, we need to be more careful. 
This is because the linear combinations of these constraints smeared against test functions may themselves be first-class. 
For this reason, we consider the class of functions of the form $E[f,g] = F[f] + G[g]$, where:
\begin{align}
    F[f] &:= \frac{i}{4 \pi^2} \int_\Sigma \! \killing{f(x)}{\varphi(z) F(A)} \, d^3x - \frac{1}{2 \pi} \sum_{p \in P \cup Z}  \int \tr{f(x)}{A_x}{p} \, d x \ca  \\
    G[g] &:= \frac{-i}{4 \pi^2} \int_\Sigma \! \killing{g(x)}{\varphi(z) G(B)} \, d^3x + \frac{1}{2 \pi} \sum_{p \in P \cup Z}  \int \tr{g(x)}{B_x}{p} \, d x \fp
\end{align} 
are the smeared Gauss laws. 
We have included the boundary terms in the above expression to ensure the functionals are differentiable. 
Importantly, the linear combination of constraints $E[f,g]$ must itself weakly vanish on the constraint surface after any boundary conditions are imposed. 
This is only the case if the pair of test functions $(f,g) \in \lie g \times \lie h$ is smooth and satisfies the boundary condition $f(q) = g(q)$ for each $q \in P$. 
We denote the class of such pairs by $I(P)$. 

Using the only non-trivial Dirac brackets between the functionals\footnote{The boundary terms in the following expressions can be thought of as a central extension of the Poisson algebra. This is a well-known phenomenon first noticed in \cite{Brown:1986ed,Brown:1986nw}.} $F[f]$ and $G[g]$: 
\begin{align}
    \{ F[f_1],F[f_2] \}_1 &= F\left[[f_1,f_2]\right] + \frac{1}{2 \pi} \sum_{p \in P \cup Z} \int \! \tr{f_1}{f_2'}{p} \, dx\ca \label{eq: F bracket} \\
    \{ G[g_1],G[g_2] \}_1 &= G\left[[g_1,g_2]\right] - \frac{1}{2 \pi} \sum_{p \in P} \int \! \tr{g_1}{g_2'}{p} \, dx \ca \label{eq: G bracket}
\end{align} 
we find that: 
\begin{align}
    \{E[f_1,g_1] \, , \, E[f_2,g_2]\} &= E[[f_1,f_2],[g_1,g_2]] + \frac{1}{2 \pi} \sum_{p \in P \cup Z} \int \left( \tr{f_1}{f_2'}{p} - \tr{g_1}{g_2'}{p} \right) \, dx \nonumber \\
    & \approx \frac{1}{2 \pi} \sum_{p \in P \cup Z} \int \left( \tr{f_1}{f_2'+[A_x,f_2]}{p} - \tr{g_1}{g_2'+[B_x,g_2]}{p} \right) \, dx \nonumber \\
    & = \frac{1}{2 \pi} \sum_{q \in P} \int \tr{g_1}{[A_x -B_x,g_2]}{q} \, dx \nonumber \\
    &= 0 \fp
\end{align} 
where in third equality we have used the fact that the pairs $(f_{1,2},g_{1,2})$ are elements of $I(P)$, and in the fourth, imposed the boundary condition $\iota_q^*A_x = \iota_q^* B_x$ after which we use $[\lie h,\lie f]\subset\lie f$ and $\killing{\lie h}{\lie f} = 0$. 
It thus immediately follows that $\varphi(z) F \approx 0$ and $\varphi(z) G \approx 0$ are first-class. 

Dirac's conjecture \cite{Dirac-1964,Henneaux-book} states that first-class constraints are generators of gauge transformations, which is certainly true of $E[f,g]$ as a short calculation shows that $\{ E[f,g], A_a \}_1 = \partial_a f + [A_a,f]$ and $\{E[f,g],B_a\}_1 = \partial_a g + [B_a,g]$ even when $(f,g) \not\in I(P)$. 
In fact, this provides a natural interpretation of the gauge transformations generated by $E[f,g]$ for $(f,g) \in I(P)$ as the class of `proper' gauge transformations, which are those that leave the boundary physics unchanged. 
The reader is likely also aware of `improper' gauge transformations, for which this is not true; we discuss these in the next section. 

As the Gauss laws are themselves first-class, we must ask if there is a set of `tertiary' constraints that follow from the requirement that they hold for all time. 
Merecifully, this is not the case as one finds that $\{E[f,g],H\} \approx 0$ from the requirement that $f$ and $g$ be elements of the class $I(P)$. 
This is just the statement that the Hamiltonian is gauge-invariant.

In the succeeding sections, we will discuss the time evolution of gauge-dependent quantities, which, as they evolve, can simultaneously change under a gauge transformation. 
This was not an issue in the preceding analysis, as we were discussing first-class constraints, which are, of course, invariant on the associated constraint surface. 
To correctly describe their dynamics, we use the `extended Hamiltonian' in which one includes linear combinations of all first-class constraints, since these are the generators of gauge transformations. 
In our scenario, the extended Hamiltonian is: 
\begin{align}
    H_E &= H + E[U_t,V_t] + \int_\Sigma \! \left( \killing{U_1}{P^t} + \killing{V_1}{ Q^t} \right) \, d^3 x \ca
\end{align} 
in which $U_1$ and $V_1$ are arbitrary functions and $U_t, V_t \in I(P)$. 

Using $H_E$, we find the equations of motion are: 
\begin{align}
    \dot{A}_a = \partial_a (A_t +U_t) + [A_a,A_t + U_t] \ca \qquad \dot{A}_t = U_1 \ \\
    \dot{B}_a = \partial_a (B_t +V_t) + [B_a,B_t + V_t] \ca \qquad \dot{B}_t = V_1 \ca
\end{align}
where $a= x,\bar{z}$. 
Notice that the dynamics of the theory are determined in the bulk up to arbitrary gauge transformations. 
This is exactly as one would expect in a generally covariant theory in which the bulk part of the Hamiltonian is always a linear combination of first-class constraints \cite{Henneaux-book}. 

\subsection{Edge Modes and Improper Gauge Transformations\label{sec:edgemodes}} 

In the next section, we will gauge fix the constraints that we have just found and reduce the theory down to the associated constraint surface. 
Before we do this, we can get an idea of what to expect by calculating the Poisson brackets for gauge-invariant observables, as these will remain unchanged by the gauge-fixing process. 
There are, of course, two obvious candidates for such observables: Wilson lines, which we will not discuss, and `edge modes' \cite{Balachandran:1991dw,Balachandran:1992yh} (gauge-invariant functionals that reduce to boundary terms once the constraints are imposed). 

To explicitly illustrate what we mean by an edge mode, suppose that $\lie h = \varnothing$, which reduces the set of proper gauge transformations down to those generated by $F[f]$ such that $f(q) = 0$ for all $q \in P$. 
The simplest possible edge mode functional is thus $F[\xi_q^{n}]$, where $\xi_q^n(x)$ is assumed to be a smooth function that $i)$ vanishes at the points $P \setminus q$, $ii)$ is the identity at the points in $Z$ and $iii)$ at $q$ is given by $\xi_q^{n}(q) = \exp \left( \frac{-inx}{R} \right) T^I $. 
Using \cref{eq: F bracket}, it is simple to show two properties of $F[\xi_q^n]$; the first being that it is gauge invariant due to the condition $f(q) = 0$. 
The second property is that on the constraint surface $P (\mathscr C_2)$, $F[\xi_q^n]$ reduces to a mode\footnote{We can similarly introduce an edge mode $A_n^{(\zeta) I} = - k_\zeta^{-1} F[\xi_\zeta^n]$ associated to the surface defect $W_\zeta$ by inserting the test function $\xi_\zeta^n = (z-\zeta)^{-1} \exp \left( -\frac{in x}{R} \right) T^I$. However, when calculating the bracket $\{ A_n^{(\zeta) I}, A_m^{(\zeta) J}\}$ we find in $\varphi(z)\langle [\xi_\zeta^n , \xi_\zeta^m] \, , \, A_x \rangle$ a double pole which needs to be regularised. We wish to avoid introducing such a procedure and thus defer discussing these modes. That being said, we expect the brackets amongst these modes to be those found from taking residues of the Maillet bracket.}: 
\begin{equation}
    F[\xi_q^{n}] \approx - \frac{k_q}{2 \pi} \int A_x^I (q) \exp \left( -\frac{in x}{R} \right) dx =: A^{(q) I}_n
\end{equation} 
of a Poisson Kac-Moody algebra at level $-k_q$: 
\begin{equation}
    \{ A^{(q)I}_n \, , \, A^{(q)J}_m \}_1 \approx f^{IJ}_{\hspace{3mm} K} A_{n+m}^{(q)K} + i n  k_q \delta_{n+m,0} C^{IJ} \ca \label{eq:edge-mode-brackets}
\end{equation}
where $C^{IJ} = \killing{T^I}{T^J}$. 
This bracket provides the first hint of the Poisson structures normally encountered in integrable models. 
To see this, we define the function: 
\begin{equation}
    L_n^I (z) = \frac{1}{\varphi(z)} \sum_{q \in P_N} \frac{A^{(q)I}_n}{z-q} \ca
\end{equation}
and then use \cref{eq:edge-mode-brackets} to calculate its bracket with itself. 
Doing this yields the mode expansion of the Maillet bracket for a Lax matrix:
\begin{equation}
    \{ L_n^I(z),L_m^I (w) \}_1 \approx f^{IJ}_{\hspace{3mm}K} \left( r(w,z) L^K_{n+m}(z) + r(z,w) L^K_{n+m}(w)\right) + i n \left( r(w,z) + r(z,w) \right) \delta_{n+m,0} C^{IJ} \ca
\end{equation}
in which $r(w,z) = \varphi(w)^{-1}(w-z)^{-1}$. 

In the previous subsection, we mentioned that the class of gauge transformations can be divided into proper and improper transformations. 
It is to this latter class that we now turn our focus. 
Unlike their proper cousins, improper transformations are not generated by first-class constraints. 
Instead, their generators are the functionals $F[\alpha(x)]$, in which the test function $\alpha(x) \notin I(P)$ satisfies any of the conditions imposed by the requirement that $F[\alpha(x)]$'s action preserve boundary conditions. 
What makes these transformations particularly interesting is that they map (but do not identify) physical states into each other. 
For example, if we assume $\alpha(x)$ is holomorphic on $\mathbb C P^1$ then the edge mode $A_{n}^{(q) I}$ transforms in the usual way under the action of $F[\alpha(x)]$: 
\begin{equation}
    \{ F[\alpha(x)] , A^{(q)I}_n \}_1 \approx - i n k_q \alpha^I_{-n} - f^{I}_{\hspace{1.5mm}JK} \sum_{m = - \infty}^\infty \alpha_m^J A_{n+m}^{(q)K} \ca
\end{equation}
in which we have used $\alpha(x) = \sum_{m} \alpha_m^I \exp \left( -\frac{inx}{R} \right) T_I$. 
Note that the above discussion did not require the use of chiral $A_x = A_t$ or anti-chiral $A_x = -A_t$ boundary conditions, and thus it holds generally. 

Suppose now we relax the condition $\lie h = \varnothing$, which simultaneously expands the set of proper gauge transformations (as $f$ in $E[f,g]$ is now determined by $g \in \lie h$) and shrinks the set of improper ones. 
This offers up an interesting interpretation for what happens when we couple the two 4dCS theories: we are gauging an $\mathcal H$ subgroup of the improper transformations. 
Given this fact, it follows that a natural strategy for constructing the edge mode observables with doubled 4dCS is to look for $\mathcal H$-invariant objects. 
To be precise, for each of the surface defects $W_q$ associated with a pole $q \in P$, we introduce a set of edge modes $A_n^{(q)I}$ in exactly the way described above\footnote{We ignore the $B_n^{(q)i}$'s because the boundary conditions mean that these are determined by projecting $A$'s modes.}. 
An infinite set of observables is then given by $\mathcal{H}$-invariant polynomials of $A_{n}^{(q)}I$ \cite{Bais:1987zk}. 
For now, we will leave developing this line of reasoning to section \ref{section:coset-ifts} and move on to fixing our first-class constraints.

\section{Gauge Fixing in the Bulk and Defect Constraints \label{sec: Bulk Gauge fixing}} 

In the previous section, we found $\mathscr C_2$ formed a complete set of first-class constraints for doubled 4dCS. 
Our goal in this section is to supplement these constraints with a series of gauge choices such that the full system of equations forms a set $\mathscr C_3$ of eight second-class constraints. 
Using these, we can reduce the phase space down to $\mathscr P (\mathscr C_3)$ and calculate the corresponding set of Dirac brackets. 
These will be found by imposing the constraints within $\mathscr C_3$ on the symplectic two-form $\Omega^{(1)}$ to find a new two-form $\Omega^{(3)}$ on $\mathscr P (\mathscr C_3)$, which we then invert as before. 

This analysis starts with an observation about which gauge choices we are allowed to pick. 
Having made this point, we will then perform the gauge fixing procedure in two steps, starting with $\varphi(z) F \approx 0$ and $P^t \approx 0$ and then finishing with $\varphi(z) G\approx 0$ and $Q^t \approx 0$. 

\subsection{Admissible Gauge Choices and Initial Set Up} 

To illustrate why we cannot choose an arbitrary set of gauge choices, we consider the Gauss law constraints $\varphi(z) F \approx 0$ and $\varphi(z) G \approx 0$. 
Naively, the simplest thing to do would be to require that $A_{\bar{z}} \approx 0$ and $B_{\bar{z}} \approx 0$, which we know from the analysis of 4dCS performed in \cite{Vicedo:2019dej} would yield two copies of Maillet's bracket \cite{Maillet:1985fn,Maillet:1985ek} on $\mathscr{P}_C$. 
Unfortunately, this choice of conditions is inappropriate as it requires that we restrict the theory's degrees of freedom, as we will now argue. 

We know from the Lagrangian analysis in \cite{Stedman:2021wrw} that the most general solution of $\varphi(z) F \approx 0$ and $\varphi(z) G \approx 0$ can be found by simultaneously gauge-fixing and performing a change of variables. 
This is done by taking\footnote{Notice that the expressions are invariant under the transformation $\hat{g} \rightarrow \hat{g} k_g$ where the group element $k_g\in \mathcal{G}$ depend on $x$ and $t$ only. 
We use this symmetry to set $\hat{g}(z=\infty) =1$. }:
\begin{equation}
    A_{\bar{z}} = -\bar{\partial} \hat{g} \hat{g}^{-1} \ca \quad A_x = - \hat{g}' \hat{g}^{-1} + \hat{g} \mathcal{A} \hat{g}^{-1} \ca \quad B_{\bar z} = 0 \label{eq: gauge choice for A and B}
\end{equation}
in which $\hat{g}$ satisfies the archipelago conditions of \cite{Delduc:2019whp}. 
For $A_{\bar{z}} \approx 0$, $B_{\bar{z}} \approx 0$ to be a valid gauge choice, we must be able to perform an admissible gauge transformation\footnote{This is a gauge transformation that preserves the boundary conditions on the fields.} into the gauge.
Starting from the configuration in \cref{eq: gauge choice for A and B}, the required gauge transformation is $A \rightarrow \hat{g}^{-1} (A + d)\hat{g}$ and $B \rightarrow B$, but this is only admissible if we impose the restriction $\hat{g} \in C ( \lie{h} ) \times N ( \lie{f} )$, which kills off physical degrees of freedom. 
We have no a priori reason to do so and therefore cannot choose $A_{\bar{z}} \approx 0$, $B_{\bar{z}} \approx 0$. 

Having observed that we cannot choose $A_{\bar z}$, what can we do? 
Instead, we work with a weaker version of \cref{eq: gauge choice for A and B} which is defined as follows\footnote{We do this because \cref{eq: gauge choice for A and B} is rather strong, and using it complicates the analysis of a series of first-class constraints that will be introduced shortly.}. 
Suppose that $C^{\infty}_{\text{Arch.}} (M, \mathcal G \times \mathcal H/\mathcal Z)$ is the set of group elements in $\mathcal G \times \mathcal H / \mathcal Z$ that satisfy the archipelago conditions of \cite{Delduc:2019whp}. 

We will assume that the elements of the class $ C^{\infty}_{\text{Arch.}} (M, \mathcal G \times \mathcal H/\mathcal Z)$ have an explicit dependence upon time, and that the Hamiltonian does not generate their dynamics. 
Given a pair $(\hat g , \hat h) \in C^{\infty}_{\text{Arch.}} (M, \mathcal G \times \mathcal H/\mathcal Z)$ we introduce the gauge choices: 
\begin{equation}
    C_F = A_{\bar z} + \bar \partial \hat g \hat g^{-1} \approx 0 \ca \qquad K_G = B_{\bar z} + \bar \partial \hat h \hat h^{-1} \approx 0 \ca
\end{equation} 
whose solutions define a submanifold of $\mathscr P \times C^{\infty}_{\text{Arch.}} (M, \mathcal G \times \mathcal H/\mathcal Z)$ on which the archipelago condition are satisfied. 
Of course, these conditions only apply at a fixed time, so we must also require that $\dot C_F \approx 0$ and $\dot K_G \approx 0$. 
We can find explicit expressions for these conditions in terms of the fields and the pair $(\hat g, \hat h)$ by using $\dot f = \partial_t f + \{ f, H \}$ (which produces the equations of motion for a function $f$ with explicit time dependence) and the equations of motion for $A_{\bar z}$ and $B_{\bar z}$. 
In particular, we find: 
\begin{align}
    \dot C_F = \bar \partial (A_t + U_t) - \partial_t (-\bar \partial \hat g \hat g^{-1} ) + [A_{\bar z},A_t + U_t] &\approx 0 \ca \\
    \dot K_G = \bar \partial (B_t + V_t) - \partial_t (-\bar \partial \hat h \hat h^{-1} ) + [B_{\bar z},B_t + V_t] &\approx 0 \fp
\end{align} 

So that our argument is clear, we will proceed in two steps. 
In step one, we will use the constraint $C_F \approx 0$ and a change of variables to solve $\varphi(z) F \approx 0$ and $\dot C_F \approx 0$. 
The solution is substituted into the symplectic two-form, and a new set of Poisson brackets is found via inversion. 
In step two, we repeat these calculations for $K_G \approx 0$, $\varphi(z) G \approx 0$ and $\dot K_G \approx 0$. 

\subsection{Step One: $\varphi(z) F\approx 0$ and $P^t \approx 0$\label{subsec:gauge-fixing-step-one}}

Starting then with step one, we impose strongly $C_F = 0$ so that $A_{\bar z} = - \bar \partial \hat g \hat g^{-1}$, perform the change of variables: 
\begin{equation}
    A_x  = - \hat{g}' \hat{g}^{-1} + \hat{g} \mathcal{A} \hat{g}^{-1} \ca
\end{equation}
and substitute these expressions for $A_{\bar z}$ and $A_x$ into $\varphi(z) F = 0$. 
This simplifies the constraint to $\bar \partial \mathcal A = 0$, implying that $\mathcal A$ is holomorphic on $\mathbb CP^1_{\omega}$ (we have dropped $\varphi(z)$ as it is non-zero everywhere on $\mathbb C P^1_{\omega}$ and so is irrelevant to satisfying the equality). 
Of course, the punctures of $\mathbb C P^1_{\omega}$ are located at the points in $P \cup Z$, and so the most general holomorphic function on $\mathbb C P^1_{\omega}$ is a meromorphic function on $\mathbb C P^1$ with poles at those locations. 
It is at this point that the boundary conditions imposed at $Z$ and $P$-type defects become relevant. 
When these were introduced, we assumed two key properties of $A$: that it was regular at $P$ and allowed to be singular at $Z$. 
Thus, the most generic solution of the constraint is: 
\begin{equation}
    \mathcal A(x,z) = \mathcal A_0(x) + \sum_{\zeta \in Z} \frac{ \mathcal A_\zeta(x) }{z-\zeta} \fp \label{eq: Lax solution}
\end{equation} 
which has an associated Gaudin matrix given by: 
\begin{equation}
    \Gamma(x,z) := \frac{\varphi(z)}{2 \pi} \mathcal A(x,z) = \sum_{q \in P_N} \frac{\Gamma_q(x)}{z-q} \ca
\end{equation} 
We will call $\mathcal A$ a Lax matrix\footnote{In the wider literature, both $\mathcal A$ and $\Gamma$ are called Lax matrices; we have chosen to give them different names to avoid any confusion.} and define $\Gamma_{\infty} := \sum_{q \in P_N} \Gamma_q$ to keep the notation in the following as compact as possible. 
Importantly, when the constraint $\varphi(z) F \approx 0$ holds strongly, the corresponding term in the Hamiltonian vanishes and the bulk dynamics of $A_{\bar z}$ is determined by $\hat g$. 

Having solved the constraint $\varphi(z) F \approx 0$ we now substitute the solution into $\Omega^{(1)}$ and use the result to find a set of Dirac brackets on the associated constraint surface. 
Focusing on the $\delta A_{\bar z} \delta A_x$ term, as it is the only one that will change, we find:  
\begin{align}
    \Omega_A :=& \frac{i}{4 \pi^2} \int_\Sigma \! \varphi(z) \killing{\delta A_{\bar{z}}}{\delta A_x} d^3 x = \frac{i}{4 \pi^2} \int_\Sigma \! \varphi(z) \killing{ \delta ( \bar{\partial} \hat{g} \hat{g}^{-1})}{\delta ( \hat{g}' \hat{g}^{-1} - \hat{g} \mathcal{A} \hat{g}^{-1} )} \, d^3 x\label{eq: A symplectic form} \\
    =& \frac{i}{4 \pi^2} \int_\Sigma \! \varphi(z) \left( \killing{\bar{\partial} ( \hat{g}^{-1} \delta \hat{g})}{\partial_x (\hat{g}^{-1} \delta \hat{g})} - \killing{\bar{\partial}(\hat{g}^{-1} \delta \hat{g} \hat{g}^{-1} \delta \hat{g})}{\mathcal{A}} - \killing{\bar{\partial}( \hat{g}^{-1} \delta \hat{g})}{\delta \mathcal{A}} \right) \, d^3 x \nonumber \\
    =& \frac{i}{4 \pi^2} \int_\Sigma \! \bar{\partial} \left( \frac{\varphi(z)}{2} \killing{ \hat{g}^{-1} \delta \hat{g}}{ \partial_x (\hat{g}^{-1} \delta \hat{g})} - \killing{\hat{g}^{-1} \delta \hat{g} \hat{g}^{-1} \delta \hat{g}}{\varphi(z) \mathcal{A}} - \killing{\hat{g}^{-1} \delta \hat{g}}{\delta (\varphi(z)\mathcal{A})} \right) \, d^3 x \nonumber
\end{align}
where in the third equality, we have expanded out the action of $\delta$ and used the identity $\delta (\partial_i \hat{g} \hat{g}^{-1} ) = \hat{g} \partial_i ( \hat{g}^{-1} \delta \hat{g}) \hat{g}^{-1}$, and in the fourth that: 
\begin{align}
     \int_\Sigma \! \varphi(z) \killing{\bar{\partial} ( \hat{g}^{-1} \delta \hat{g})}{\partial_x (\hat{g}^{-1} \delta \hat{g})} \, d^3 x &= \frac{1}{2} \int_\Sigma \! \varphi(z) \left( \killing{\bar{\partial} ( \hat{g}^{-1} \delta \hat{g})}{\partial_x (\hat{g}^{-1} \delta \hat{g})} - \killing{\partial_x \bar{\partial} ( \hat{g}^{-1} \delta \hat{g})}{\hat{g}^{-1} \delta \hat{g}} \right) \, d^3 x\nonumber \\
    &= \frac{1}{2} \int_\Sigma \! \varphi(z) \bar{\partial} \killing{\hat{g}^{-1} \delta \hat{g}}{\partial_x (\hat{g}^{-1} \delta \hat{g})} \, d^3 x\ca
\end{align}
and $\bar{\partial} (\varphi (z) \mathcal A) = 0$ on $\mathbb C P^1_\omega$. 
Using \cref{eq:boundary-integral-result} we can localise \cref{eq: A symplectic form} down to the surface defects, which, after using the definition of the Gaudin matrix, becomes $\Omega_A = \sum_{q \in P_N} \Omega_q$ where\footnote{The sum in $\Omega_A$ is performed over $P_N$ because our choice to define $\hat{g}$ such that $g_{\infty} =1$ implies $\Omega_{\infty} = 0^A$.}: 
\begin{align}
    \Omega_q^A = \int \left( \frac{k_q}{4\pi} \killing{g_q^{-1} \delta g_q}{ \partial_x (g^{-1}_q \delta g_q )} - \killing{g_q^{-1} \delta g_q g^{-1}_q \delta g}{\Gamma_q} - \killing{g^{-1}_q \delta g_q}{\delta \Gamma_q} \right) \, dx \ca \label{eq:Initial-A-defect-two-form}
\end{align} 
with $g_q = \iota_q^*\hat{g}$ . 
There are no contributions to $\Omega$ from the punctures located at the $\omega$'s zeros because $\hat{g}$ is regular at these locations. 

In the final step of this calculation, we introduce a set of group coordinates\footnote{There is a separate set of coordinates defined for each $q \in P$, we just suppress this fact in our notation to ensure it does not become too cumbersome.} $\{\theta^a\}$ on $\mathcal{G}$ and define a matrix $J_a^I$ by:
\begin{equation}
    J_a^I T_I = \frac{k_q}{2 \pi} g^{-1}_q \partial_a g_q = J_a \ca
\end{equation} 
where $\partial_a$ denotes the derivative with respect to $\theta^a$. 
As it will be momentarily useful, we also define the inverse $j^{a}_{\, J}(x)$ of $J_a^I$, i.e.\ $J_a^I(x) j^{a}_{\, J}(x) = \delta^I_{\, J}$, and observe that $J_a$ satisfies the flatness condition $\partial_a J_b - \partial_b J_a + \tfrac{2 \pi}{k_q}[J_a,J_b] = 0$. 
The above equation enables us to set $g^{-1}_q \delta g_q = \tfrac{2 \pi}{k_q}J_a \delta \theta^a$ and thus rewrite the defect two-form as: 
\begin{align}
    \Omega_q^A &= \frac{1}{2} \frac{2 \pi}{k_q} \int \left(  \killing{J_a \delta \theta^a}{\partial_x( J_b \delta \theta^b)}
    + \tfrac{2 \pi}{k_q}\killing{J_a}{[\Gamma_q, J_b]} \delta \theta^a \delta \theta^b - 2 \killing{J_a}{T_I} \delta \theta^a \delta \Gamma^I \right) \, d x\nonumber \\
    &= \frac{1}{2} \frac{2 \pi}{k_q}\int \int \left( \langle J_a(x) \, , \, J_b(y) \delta'(x-y) + \tfrac{2 \pi}{k_q}[\Gamma_q(x), J_b(y)]\delta(x-y) \rangle \delta \theta^a(x) \delta \theta^b(y) \right. \\
    & \qquad \qquad - \delta (x-y) \killing{J_a(x)}{T_I} \delta \theta^a(x) \delta \Gamma^I_q(y) + \delta (x-y) \killing{T_I}{J_a(y)} \delta \Gamma^I_q(x) \delta \theta^a (y) \Big) \, dx dy \fp \nonumber 
\end{align} 
which, when expressed as a matrix, is made up of the blocks:
\begin{equation}
    (\Omega_{q}^A)_{IJ} = \frac{2 \pi}{k_q} \begin{pmatrix}
        \left(C_{IJ} \delta'(x-y) - \dfrac{2 \pi}{k_q}f_{IJK} \Gamma_q^K(x) \delta(x-y) \right) J_a^I(x) J_b^J(y) \ca & - \delta(x-y) J_{aI}(x) \\
        \delta(x-y) J_{bI} (y) \ca & 0
    \end{pmatrix} \ca
\end{equation}
where $C_{IJ}=\killing{T_I}{T_J}$ and whose inverse is: 
\begin{equation}
    \begin{pmatrix}
        0 \ca& \dfrac{k_q}{2 \pi} j^{bI}(y) \delta(y-x') \\
       - \dfrac{k_q}{2 \pi} j^{bI}(x') \delta(y-x') \ca & \dfrac{k_q}{2 \pi} C^{IJ} \delta' (y-x') - f^{I J}_{\quad \! K} \Gamma_q^K(y) \delta (y-x') 
    \end{pmatrix} \fp
\end{equation}
Thus, after dressing the Lie algebra indices with the basis elements of $\lie{g}$, we find the non-zero Poisson brackets are a Kac-Moody current algebra at level $k_q$: 
\begin{align}
    \{ \Gamma_{\! q \textbf{1}}(x) , \Gamma_{\! q \textbf{2}}(y) \} = [ \mathsf{C} , \Gamma_{\! q \textbf{1}}(x)] \delta(x-y) - \frac{k_q}{2 \pi} \, \mathsf{C} \, \delta'(x-y) \ca \quad \{ \Gamma_{\! q}(x), \theta^a(y) \} = \frac{k_q}{2 \pi} j^{a} (y) \delta(x-y) \ca \label{eq: fundamental brackets 1} 
\end{align}
justifying our earlier choice to call the coefficients of $\omega$ levels. 
Using the above equations alongside $(w-z)(z-q)^{-1}(w-q)^{-1} = (z-q)^{-1}-(w-q)^{-1}$, and $[\, \mathsf{C} \, , \Gamma_\textbf{1} (x,z)] = -[\mathsf{C}, \Gamma_\textbf{2}(x,z)]$, followed by an integration against test functions, one finds the Maillet-Gaudin bracket: 
\begin{equation}
    \{\Gamma_\textbf{1}(x,z), \Gamma_\textbf{2}(y,w) \} = \left[\frac{\mathsf{C} \, \delta(x-y)}{w-z},\Gamma_\textbf{1}(x,z) + \frac{\varphi(z)}{2 \pi} \partial_x \right] - \left[\frac{ \mathsf{C} \,  \delta(x-y)}{z-w},\Gamma_\textbf{2}(y,w) + \frac{\varphi(w)}{2 \pi} \partial_y \right] \ca
\end{equation} 
and thus that $\mathcal{A}(x,z)$ satisfy the Maillet bracket: 
\begin{equation}
    \{ \mathcal{A}_{\textbf{1}}(x,z), \mathcal{A}_\textbf{2}(y,w) \} = \left[ \mathsf{R}(w,z) \delta(x-y),\mathcal{A}_\textbf{1}(x,z) + \partial_x \right] - \left[\mathsf{R}(z,w) \delta(x-y),\mathcal{A}_\textbf{2}(y,w) + \partial_y \right] \ca
\end{equation} 
where we call: 
\begin{equation}
    \mathsf{R}(w,z) := \frac{2 \pi}{\varphi(w)} \frac{\mathsf{C}}{w-z} \ca
\end{equation}
an $r$-matrix. 
Since it will be useful later on, we note that: 
\begin{align}
    \{ \Gamma_{\! q \textbf{1}} (x), g_{q \textbf{2}} (y) \} &= \{ \Gamma_{\! q \textbf{1}} (x), \theta^a(y) \} \partial_a  g_{q \textbf{2}} (y) = g_{q \textbf{2}} (y)   \mathsf C \delta (x-y) \label{eq:Gamma-g-bracket}
\end{align} 
which follows from $g_q J_{a} = k_q \partial_a g_q / 2 \pi$ and the definition of $j^a(y)$. 

Let us turn now to solving $\dot C_F \approx 0$, which simplifies to: 
\begin{equation}
    \bar \partial \left ( \hat g^{-1} \partial_t \hat g + \hat g^{-1} (A_t + U_t) \hat g \right) \approx 0 \label{eq:A_t-constraint}
\end{equation} 
as $C_F = 0$ has been imposed strongly. 
Importantly, this constraint does not contain $\mathcal A$ or the defect fields $\{ g_q\}$ and so its imposition does not change the brackets just found above. 
By carefully taking into account the boundary conditions required of $A_t$ and $U_t$ at the points $Z$, we can solve this constraint and find that: 
\begin{equation}
    A_t + U_t = - \partial_t \hat g \hat g^{-1} + \hat g \mathcal M \hat g^{-1} \ca \quad \text{where} \quad \mathcal M = \mathcal M_0 + \sum_{\zeta \in Z} \frac{\alpha_\zeta \mathcal A_\zeta}{z-\zeta} \ca
\end{equation}
which completely determines $A_t$ in terms of $\hat g$ and $\mathcal A$, up to the holomorphic part of $\mathcal M$ which will be fixed by using the boundary condition at $\infty$ momentarily. 

\subsection{Step Two: $\varphi(z) G \approx 0$ and $Q^t \approx 0$} 

We can now repeat this analysis for $\varphi(z) G \approx 0$ by using the gauge choice $K_G \approx 0$ and the change of variables:
\begin{align} 
    B_x = - \hat{h}' \hat{h}^{-1} + \hat{h} \mathcal{B} \hat{h}^{-1} \fp \label{eq: change of variables B}
\end{align} 
The resulting series of calculations is substantively the same as those we have just done and produces the solutions: 
\begin{equation}
    \mathcal  B(x,z) = \mathcal B_0 (z,x) + \sum_{\zeta \in Z} \frac{\mathcal B_\zeta (x)}{z-\zeta} \ca 
\end{equation}
which also has an associated Gaudin matrix: 
\begin{equation}
    \Delta(x,z) := \frac{\varphi(z)}{2 \pi} \mathcal B(x,z) = \sum_{q \in P_N} \frac{\Delta_q(x)}{z-q} \fp
\end{equation} 
In the next step, we insert into $\Omega^{(2)}$ $B_{\bar z} = - \bar \partial \hat h \hat h^{-1}$, \cref{eq: change of variables B} and the above solution for $\mathcal B$ to find a boundary term which is then collapsed down to a sum over the poles $P$ by using \cref{eq:boundary-integral-result}. 
Finally, we introduce a set of group coordinates $\{ \sigma^a \}$ on $\mathcal H$ and define a matrix $I_a^i$ by $I_a = \tfrac{k_q}{2 \pi} h_q^{-1} \partial_a h_q$ which is again flat and whose inverse is $i^{a}_{\ j}(x)$. 
In the end, we find that the $\delta B_{\bar z} \delta B_x$ component of the symplectic two-form $\Omega^{(2)}$ is $\Omega^B = \sum_q \Omega^B_q$, where:
\begin{align}
    \Omega_q^B &= -\frac{1}{2} \frac{2 \pi}{k_q}\int \int \left( \langle I_a(x) \, , \, k_q I_b(y) \delta'(x-y) + \tfrac{2 \pi}{k_q}[\Delta_q(x), I_b(y)]\delta(x-y) \rangle  \delta \sigma^a(x) \delta \sigma^b(y) \right. \\
    & \qquad \qquad - \delta (x-y) \killing{I_a(x)}{T_i} \delta \sigma^a(x) \delta \Delta^i_q(y) + \delta (x-y) \killing{T_i}{I_a(y)} \delta \Delta^i_q(x) \delta \sigma^a (y) \Big) \,  d x  dy \fp \nonumber 
\end{align} 
which, upon inverting as a matrix, gives the Poisson brackets: 
\begin{equation}
    \{ \Delta_{q \textbf{1}}(x) , \Delta_{q \textbf{2}}(y) \} = \frac{k_q}{2 \pi} \, \mathsf{C} \, \delta'(x-y) - [\, \mathsf{C} \,, \Delta_{q \textbf{1}}(x)] \delta(x-y) \ca \quad \{ \Delta_{q}(x), \sigma^a(y) \} = - \frac{k_q}{2 \pi}i^{a} (y) \delta(x-y) \ca \label{eq: fundamental brackets 2}
\end{equation} 
Therefore, we find a second copy of both the Maillet and Maillet-Gaudin brackets: 
\begin{align}
    \{\Delta_\textbf{1}(x,z), \Delta_\textbf{2}(y,w) \} &= \left[\frac{\mathsf{C}^{\lie{h}}\delta(x-y)}{z-w} ,\Delta_\textbf{2}(y,w) + \frac{\varphi(w)}{2 \pi} \partial_y \right] - \left[\frac{\, \mathsf{C}^{\lie{h}} \delta(x-y)\,}{w-z},\Delta_\textbf{1}(x,z) + \frac{\varphi(z)}{2 \pi} \partial_x \right] \ca \\
    \{ \mathcal{B}_{\textbf{1}}(x,z), \mathcal{B}_\textbf{2}(y,w) \} &= [\mathsf{R}^{\lie{h}}(z,w) \delta(x-y),\mathcal{B}_\textbf{2}(y,w) + \partial_y] -[\mathsf{R}^{\lie{h}}(w,z) \delta(x-y),\mathcal{B}_\textbf{1}(x,z) + \partial_x] \fp 
\end{align}
The sign difference in these brackets when compared to those for $\mathcal A$ and $\Gamma$ is explained by the relative difference in the initial action between the $CS(A)$ and $CS(B)$ terms. 
By following the same argument used to calculate $\{ \Gamma_{\! q \textbf{1}} (x) , g_{q \textbf{2}} (y) \}$ we find that: 
\begin{equation}
    \{ \Delta_{q \textbf{1}} (x) , h_{q \textbf{2}} (y) \} = - h_{q \textbf{2}} (y) \mathsf{C} \delta (x-y) \fp 
\end{equation} 

Having fixed the constraint $\varphi(z) G \approx 0$ using $K_G \approx 0$ we now move on to fixing $P^t \approx 0$. 
To do this we use $\dot K_G \approx 0$ which, after $K_G \approx 0$ is imposed, becomes: 
\begin{equation}
    \bar \partial \left ( \hat h^{-1} \partial_t \hat h + \hat h^{-1} (B_t + V_t) \hat h \right) \approx 0 \fp \label{B_t-constraint}
\end{equation}
We can solve this constraint in a similar fashion to $\dot C_F \approx 0$ from which we find: 
\begin{equation}
    B_t + V_t = - \partial_t \hat g \hat g^{-1} + \hat g \mathcal N \hat g^{-1} \ca \quad \text{where} \quad \mathcal N = \mathcal N_0 + \sum_{\zeta \in Z} \frac{\beta_\zeta \mathcal B_\zeta}{z-\zeta} \fp
\end{equation}
Equation \eqref{B_t-constraint} does not contain $\mathcal B$ or $\{ h_q \}$ and thus its imposition does not change their brackets, just as imposing \cref{eq:A_t-constraint} did not change the brackets for $\mathcal A$ and $\{ g_q \}$. 
Having now determined $B_t$, we can use the boundary condition at infinity $A_t = \epsilon_{\infty} A_x^{\lie f} + B_t$ to fix the holomorphic component $\mathcal M_0$ of $\mathcal M$. 
When doing this we find that $\mathcal M_0 = \epsilon_{\infty} \mathcal A_0^{\lie f} + \mathcal N_0$ as $g_\infty = h_\infty =1$. 

Before moving on, we would like to make one final comment concerning $\mathcal B_0$ and $\mathcal N_0$. 
Although they essentially arise from the constraint equations as constants of integration, they play an important role in the boundary theory as the components of a flat gauge field. 
Of course, our goal at the end of this process is to find a reduced Hamiltonian in which the gauge theory is completely fixed. 
This will be achieved later on by setting $\mathcal N_0 = 0$ and $\mathcal B_0' = 0$. 

\section{The Boundary Hamiltonian Analysis\label{section:boundary-Hamiltonian-Analysis}} 

This section has three purposes. 
Firstly, it aims to clarify the role of the boundary condition $A_x^{\lie h}(x,q) - B_x (x,q)= 0$ now that the bulk variables have been eliminated. 
We shall argue that these conditions should be recast as a set of `boundary constraints' on the variables associated with each defect. 
This then brings us to the section's second purpose, which is to perform the Dirac analysis of these constraints and eliminate all but one of them. 
We then conclude the section by giving a rudimentary definition of a new Poisson algebra associated to the class of integrable models produced by doubled 4dCS. 

\subsection{The Reduced Hamiltonian and Defect Constraints} 

When we were gauge fixing and imposing constraints in the previous section, we could have simultaneously imposed some of the boundary conditions located at the poles $P_N$. 
Doing so would allow us to express $\Gamma (x,z)$ in terms of the $g_q(x)$'s and $\Delta_q(x)$'s. 
We know from the Lagrangian analysis performed in \cite{Stedman:2021wrw}, that doing so produces the Lax matrix for a particular gauged $\sigma$-model. 
The reason we haven't done this is that we want the substance of our analysis to hold for a generic system. 
We will explain how to find the Lax matrix at the end of section \ref{subsection:gauge fixing the defect constraints}. 

Our starting point for this argument is the assumption that only the Gauss law constraints have been strongly imposed, and thus that we have replaced the bulk variables $A_{a}(x)$ and $B_{a}(x)$ for $a = x , \bar z$ with a set of `defect degrees of freedom' $(g_q(x), h_q(x))$ and $(\Gamma_q(x), \Delta_q(x))$ for each $q \in P$. 
The dynamics of these remaining variables are generated by the reduced Hamiltonian: 
\begin{align}
    &H = \frac{1}{2} \sum_{q \in P} \int \left( \langle \iota_q^*A_t^\lie h + \iota_q^*B_t +  2 \iota_q^*V_t \, , \, ( g_q ( \Gamma_q - J_q ) g_q^{-1})^{\lie{h}} - h_q ( \Delta_q - I_q ) h_q^{-1} \rangle\right. \nonumber \\
    &\hspace{3cm}+ \left. \frac{2 \pi \epsilon_q}{k_q} \langle ( g_q \Gamma_q g_q^{-1} - g_q J_q g_q^{-1})^{\lie{f}} \, , \, ( g_q \Gamma_q g_q^{-1} - g_q J_q g_q^{-1})^{\lie{f}} \rangle \right) \, dx \\ 
    &+ \pi \sum_{\zeta \in Z} \int \left( \alpha_\zeta \bigtr{ \varphi(z)^{-2} \Gamma }{ \Gamma }{\zeta} - \beta_\zeta \bigtr{ \varphi(z)^{-2} \Delta}{\Delta}{\zeta}  \right) \, dx + \int_{\Sigma} \! \left( \killing{U_1}{P^t} + \killing{V_1}{Q^t} \right) \, d^3x \nonumber
\end{align} 
which can be found by setting $\varphi(z) F = 0$ and $\varphi(z) G = 0$, and where: 
\begin{equation}
    J_q(x) = \frac{k_q}{2 \pi} g^{-1}_q (x)g_q' (x) \ca \qquad I_q (x)= \frac{k_q}{2 \pi} h_q^{-1} (x) h_q' (x) \fp
\end{equation}

The defect degrees of freedom are `local' in the sense that they have well-defined Poisson brackets amongst each other, and thus any functional constructed from them is differentiable. 
This fact alters both the set of boundary conditions required for $H$ to be functionally differentiable and how we interpret them. 
To see this, we perform the variation: 
\begin{equation}
    \delta H = \frac{1}{2} \sum_{q \in P} \int \left( \langle \iota_q^*\delta A_t^\lie h + \iota_q^*\delta B_t \, , \, ( g_q ( \Gamma_q - J_q ) g_q^{-1})^{\lie{h}} - h_q ( \Delta_q - I_q ) h_q^{-1} \rangle\right) dx + \, \cdots \ca \label{eq:var-reduced-ham}
\end{equation}
where the terms contained in $\cdots$ are functionally differentiable and so can be ignored. 
Looking at \cref{eq:var-reduced-ham}, we can immediately see that $H$ is only differentiable if for each $q \in P_N$ we require: 
\begin{equation}
    ( g_q (x)( \Gamma_q (x) - J_q (x) ) g_q^{-1}(x))^{\lie{h}} - h_q(x) ( \Delta_q(x) - I_q (x)) h_q^{-1}(x) = 0 \ca \label{eq:dbc1}
\end{equation} 
while the term at $z=\infty$ produces: 
\begin{equation}
    \sum_{q \in P_N} (\Gamma_q^\lie h (x) - \Delta_q(x)) = 0 \fp \label{eq:dbc2}
\end{equation}
Both of the above equations are just the boundary condition $A_x^\lie h (q) = B_x (q)$ expressed in terms of the new variables. 

Unusually, these equations impose restrictions on the defect fields across the whole of the domain in which they are defined, i.e. the surface defects $W_q$. 
This is in contrast to standard boundary conditions, which restrict the fields on a subdomain. 
We have already encountered a class of functionals with this property: the constraints on the fields discussed above. 
It is for this reason that our interpretation of \cref{eq:dbc1,eq:dbc2} changes, such that we instead consider them as `defect constraints'\footnote{An alternative way of introducing $C_q \approx 0$ and $C_{\infty} \approx 0$ is as the secondary constraints that ensure $P^t \approx 0$ and $Q^t \approx 0$ hold for all time after $\varphi(x) F \approx 0$ and $\varphi(z) G \approx 0$ are eliminated.}: 
\begin{align}
    C_{q} &= ( g_q (x)( \Gamma_q (x) - J_q (x) ) g_q^{-1}(x))^{\lie{h}} - h_q(x) ( \Delta_q(x) - I_q (x)) h_q^{-1}(x) \approx 0 \ca \\
    C_{\infty} &= \sum_{q \in P_N} (\Gamma_q^\lie h (x) - \Delta_q(x)) \approx 0 \fp
\end{align} 

\subsection{Defect Constraint Analysis} 

In this subsection, we perform the Dirac analysis of the defect constraints and show that there are no secondary defect constraints. 
These calculations involve computing the brackets between quantities that are projected into $\lie h$. 
To do this, we use the following formulae\footnote{The subscript $\textbf{i}$ on $\killing{\cdot}{\cdot}_{\textbf{i}}$ indicates that the bilinear form is taken over the $i$-th factor in the tensor product.}: 
\begin{equation}
    \{ X^{\lie h}_{\textbf{1}} (x) , Y (y)_{\textbf{2}} \} = \langle \{ X_{\textbf{3}} (x) , Y_{\textbf{4}} (y) \} \, , \mathsf{C}_{\textbf{13}}^{\lie h} \rangle_{\textbf{3}} \ca \quad \{ X^{\lie h}_{\textbf{1}} (x) \, , \, Y^{\lie h}(y)_{\textbf{2}} \} = \langle \langle \{ X_{\textbf{3}} (x) , Y_{\textbf{4}} (y) \} \, , \mathsf{C}_{\textbf{13}}^{\lie h} \rangle_{\textbf{3}} \, , \mathsf{C}_{\textbf{24}}^{\lie h} \rangle_{\textbf{4}} \label{eq:project-brackets}
\end{equation} 
which both follow from the identity $X^{\lie h}_{\textbf{1}} = \langle X_{\textbf{2}} \, , \, \mathsf{C}_{\textbf{12}}^{\lie h} \rangle_{\textbf{2}}$. 

The subsequent analysis can be greatly simplified by first computing the Poisson brackets amongst the following quantities: 
\begin{align}
    \Gamma_{\! q}^g := g_q(x) \Gamma_{\! q} (x) g^{-1}_q(x) \ca \, \qquad J^{g}_q := g_q(x) J_q(x) g_q^{-1}(x) =  \frac{k_q}{2 \pi} g_q'(x) g_q^{-1}(x) \ca \, \\
    \Delta_q^h := h_q(x) \Delta_{\! q} (x) h^{-1}_q(x) \ca \qquad I_q^h := h_q(x) I_q(x) h_q^{-1}(x) = \frac{k_q}{2 \pi} h_q'(x) h_q^{-1}(x) \ca
\end{align} 
as well as $\{ \Gamma_{q \textbf{1}}(x), \Gamma_{q \textbf{2}}^g(y) - J_{q \textbf{2}}^g (y) \}$ and  $\{ \Delta_{q \textbf{1}}(x), \Delta_{q \textbf{2}}^h(y) - I_{q \textbf{2}}^h (y) \}$. 
Of the ten potential brackets, only the following four are non-trivial: 
\begin{equation}
    \{ \Gamma_{\! q \textbf{1}}^{g} (x) , J_{q \textbf{2}}^{g} (y) \}  \ca \quad \{ \Gamma_{\! q \textbf{1}}^{g} (x) , \Gamma_{\! q \textbf{2}}^{g} (y) \} \ca \quad \{ \Delta_{q \textbf{1}}^{h} (x) , I_{q \textbf{2}}^{h} (y) \} \ca \quad \{ \Delta_{q \textbf{1}}^{h} (x)  , \Delta_{q \textbf{2}}^{h} (y)  \} \fp 
\end{equation}
The first of these is the simplest to calculate:
\begin{align}
    \{ \Gamma_{\! q \textbf{1}}^{g} (x) , J_{q \textbf{2}}^{g} (y) \} &= g_{q \textbf{1}} (x) \{ \Gamma_{\! q \textbf{1}} (x), J_{q \textbf{2}}^{g} (y) \} g_{q \textbf{1}}^{-1} (x) \nonumber \\
    &= \frac{k_q}{2 \pi} g_{q \textbf{1}} (x) \left( \partial_y \{ \Gamma_{\! q \textbf{1}} (x), g_{q \textbf{2}} (y) \} g_{q \textbf{2}}^{-1} (y) + g_{q \textbf{2}}'(y) \{ \Gamma_{q \textbf{1}}(x) , g_{q \textbf{2}}^{-1}(y) \} \right) g_{q \textbf{1}}^{-1} (x) \nonumber \\
    &=\frac{k_q}{2 \pi} g_{q \textbf{1}} (x) g_{q \textbf{2}} (y) \mathsf{C} g_{q\textbf{1}}^{-1}(x) g_{q\textbf{2}}^{-1}(y) \delta'(x-y) \nonumber \\
    &=- \frac{k_q}{2 \pi} \mathsf{C} \delta'(x-y) + [ \mathsf{C} , J_{q \textbf{1}}^{g} (x) ] \delta(x-y) \label{eq:gamma-g-J-g-bracket}
\end{align} 
where in the fourth equality we have integrated against a test function, and then simplified the result by applying the identity $g_{\textbf{1}} g_{\textbf{2}} \mathsf{C} g_{\textbf{1}}^{-1} g_{\textbf{2}}^{-1} = \mathsf{C}$. 

The second bracket, $\{ \Gamma_{\! q \textbf{1}}^{g} (x) , \Gamma_{\! q \textbf{2}}^{g} (y) \}$, is computed in three steps. 
In the first, we expand it out into: 
\begin{align}
    \{ \Gamma_{\! q \textbf{1}}^{g} (x) , \Gamma_{\! q \textbf{2}}^{g} (y) \} =&\,  g_{q \textbf{1}}(x) g_{q \textbf{2}} (y) \{ \Gamma_{\! q \textbf{1}} (x) ,  \Gamma_{\! q \textbf{2}} (y) \} g_{q \textbf{1}}^{-1}(x) g_{q \textbf{2}}^{-1} (y) \label{eq:adg-gamma-squared-bracket} \\ 
    & \hspace{-2cm} + g_{q \textbf{2}}(y) \{ g_{q \textbf{1}}(x) , \Gamma_{\! q \textbf{2}} (y) \} \Gamma_{\! q \textbf{1}} (x) g_{q \textbf{1}}^{-1} (x) g_{q \textbf{2}}^{-1} (y) \nonumber + g_{q \textbf{1}}(x) \{ \Gamma_{\! q \textbf{1}} (x) , g_{q \textbf{2}}(y) \} g_{q \textbf{1}}^{-1} (x)\Gamma_{\! q \textbf{2}} (y) g_{q \textbf{2}}^{-1} (y) \\
    & \hspace{-2cm} + g_{q \textbf{1}}(x)  g_{q \textbf{2}}(y) \Gamma_{\! q \textbf{1}} (x) \{ g_{q \textbf{1}}^{-1} (x) , \Gamma_{\! q \textbf{2}} (y) \}g_{q \textbf{2}}^{-1} (y) + g_{q \textbf{1}}(x)  g_{q \textbf{2}}(y) \Gamma_{\! q \textbf{2}} (y) \{ \Gamma_{\! q \textbf{1}} (x) , g_{q \textbf{2}}^{-1} (y) \}g_{q \textbf{2}}^{-1} (y) \nonumber \fp
\end{align} 
In the second, we determine the first term by substituting in the bracket for $\{ \Gamma_{\! q \textbf{1}} (x) ,  \Gamma_{\! q \textbf{2}} (y) \}$, after which we integrate against a test function and apply the identity $g_{\textbf{1}} g_{\textbf{2}} \mathsf{C} g_{\textbf{1}}^{-1} g_{\textbf{2}}^{-1} = \mathsf{C}$. 
This yields: 
\begin{equation}  
     g_{q \textbf{1}}(x) g_{q \textbf{2}} (y) \{ \Gamma_{\! q \textbf{1}} (x) ,  \Gamma_{\! q \textbf{2}} (y) \} g_{q \textbf{1}}^{-1}(x) g_{q \textbf{2}}^{-1} (y) = [ \mathsf{C} , \Gamma_{\! q \textbf{1}}^{g} (x) - J_{q \textbf{1}}^{g} (x)] \delta (x-y) - \frac{k_q}{2 \pi} \mathsf{C} \delta'(x-y) \fp
\end{equation} 
In this final step, we evaluate the final four terms in \cref{eq:adg-gamma-squared-bracket}, whose sum we denote by $M_{\textbf{12}}(x,y)$. 
To do this we first use \cref{eq:Gamma-g-bracket} and then integrate against a test function, after which $g_{\textbf{1}} g_{\textbf{2}} \mathsf{C} g_{\textbf{1}}^{-1} g_{\textbf{2}}^{-1} = \mathsf{C}$ is again applied. 
This produces: 
\begin{equation}
        M_{\textbf{12}} (x,y) = [C, \Gamma_{\! q \textbf{2}}^{g} (y) - \Gamma_{\! q \textbf{1}}^{g} (x) ] \delta(x-y) \fp
\end{equation} 
and thus we find:
\begin{equation}
    \{ \Gamma_{\! q \textbf{1}}^{g} (x) , \Gamma_{\! q \textbf{2}}^{g} (y) \} = [ \mathsf C , \Gamma_{\! q \textbf{2}}^{\lie g} (y) - J_{q \textbf{1}}^{g} (x)] \delta (x-y) - \frac{k_q}{2 \pi} \mathsf C \delta' (x-u) \fp \label{eq:gamma-g-Gamma-g-bracket}
\end{equation} 

The Poisson bracket $\{ \Gamma_{q \textbf{1}}(x), \Gamma_{q \textbf{2}}^g(y) - J_{q \textbf{2}}^g (y) \}$ can be found in a similar fashion. 
In particular, we find: 
\begin{align}
    \{\Gamma_{\! q \textbf{1}}(x) , \Gamma^{g}_{q \textbf{2}} (y) -J^{g}_{q \textbf{2}}(y) \} &= g_{q \textbf{2}} (y) \left( [\mathsf C , \Gamma_{\! q \textbf{1}} (x) + \Gamma_{\! q \textbf{2}}(y)] \delta(x-y) - \frac{k_q}{2 \pi} \mathsf C\delta'(x-y) \right) g_{q \textbf{2}}^{-1}(y) \nonumber \\ 
    & \hspace{2 cm} + \frac{k_q}{2 \pi} g_{q \textbf{2}} (y) \mathsf C g_{q \textbf{2}}^{-1} (y) \delta'(x-y) = 0 \label{eq:gammaq-gammaq+current-bracket}
\end{align}
where the final equality follows after integrating a test function and using the identity $[\mathsf C , X_\textbf{1} + X_{\textbf{2}} ] = 0$, which holds for any $X \in \lie g$. 

The final three Poisson brackets can be found from those above by performing the map $(\Gamma_q, g_q , \mathsf{C}) \rightarrow (\Delta_q, h_q,  - \mathsf{C}^{\lie h})$, since this takes the brackets for $(\Gamma_q,g_q)$ to those of $(\Delta_q, h_q)$. 
Upon doing this, we find that: 
\begin{align}
    \{ \Delta_{q \textbf{1}}^{h} (x) , I_{q \textbf{2}}^{h} (y) \} &= \frac{k_q}{2 \pi} \mathsf{C}^{\lie h} \delta'(x-y) - [ \mathsf{C}^{\lie h} , I_{q \textbf{1}}^{h} (x) ] \delta(x-y) \ca \\
    \{ \Delta_{q \textbf{1}}^{h} (x)  , \Delta_{q \textbf{2}}^{h} (y)  \} &= \frac{k_q}{2 \pi} \mathsf C^{\lie h} \delta' (x-u) -[ \mathsf C^{\lie h} , \Delta_{q \textbf{2}}^h (y) - J_{q \textbf{1}}^{h} (x)] \delta (x-y)  \ca \\
    & \hspace{0cm} \{ \Delta_{q \textbf{1}}(x), \Delta_{q \textbf{2}}^h(y) - I_{q \textbf{2}}^h (y) \} =0 \fp \label{eq:deltq-deltaq+current-bracket}
\end{align} 

Using the above results and the formulae in \cref{eq:project-brackets}, it is simple to calculate the brackets between the constraints. 
We find that the only non-trivial ones are: 
\begin{equation}
    \{ C_{q \textbf{1}} (x) , C_{q \textbf{2}} (y) \} = - [ \mathsf{C}^{\lie h}, C_{q \textbf{1}}(x)] \delta(x-y)\ca \qquad \{ C_{\infty \textbf{1}} (x) , C_{\infty \textbf{2}} (y) \} = [ \mathsf{C}^{\lie h} , C_{\infty \textbf{1}} (x)] \delta(x-y) \ca
\end{equation}
which all vanish on the associated constraint surface, implying that the constraints are all first-class. 

Following Dirac's algorithm, we must ask whether the conditions $\dot C_q \approx 0$ and $\dot C_{\infty} \approx 0$ introduce further secondary constraints. 
For each constraint $C_q \approx 0$, the only term in the reduced Hamiltonian that could generate its dynamics is: 
\begin{equation}
    H_q^{\lie f} = \frac{\pi \epsilon_q}{k_q}  \int \langle ( g_q \Gamma_q g_q^{-1} - g_q J_q g_q^{-1})^{\lie{f}} \, , \, ( g_q \Gamma_q g_q^{-1} - g_q J_q g_q^{-1})^{\lie{f}} \rangle \, dx \ca
\end{equation} 
since the Poisson bracket of $C_q$ with the other terms vanishes on the constraint surface. 
Using the identities $X^\lie{h}_\textbf{1} =\killing{X_\textbf{2}}{\mathsf C^\lie h}_\textbf{2}$ and $X^\lie{f}_\textbf{1} =\killing{X_\textbf{2}}{\mathsf C^\lie f}_\textbf{2}$, as well as the brackets just calculated, we find: 
\begin{align}
    \{ C_{q \textbf{1}} (x), ( \Gamma_{\! q \textbf{2}}^g(y) -J_{q \textbf{2}}^g(y) )^{\lie{f}} \} &= \big\langle \big\langle \tfrac{k_q}{2 \pi} \delta'(x-y) \mathsf{C}_{\textbf{34}} - [ \mathsf{C}_{\textbf{34}} , \Gamma_{\! q \textbf{3}}^g(x) -J_{q\textbf{3}}^g (x) ] \delta(x-y) \, , \, \mathsf C_\textbf{13}^{\lie h}\big\rangle_{\textbf{3}} \, , \, \mathsf{C}^{\lie f}_{\textbf{24}} \big\rangle_{\textbf{4}} \nonumber \\ 
    &= [\mathsf{C}^\lie h , ( \Gamma_{\! q \textbf{2}}^g (x) - J_{q \textbf{2}}^g (x) )^\lie f ] \delta(x-y) \ca
\end{align}
where the final equality follows after using $[\mathsf C_{\textbf{34}} , X_\textbf{3} + X_{\textbf{4}} ] = 0$, and applying the properties $[\lie h, \lie h] \subseteq \lie h$, $[\lie h , \lie f] \subset \lie f$, and $\killing{\lie f}{ \lie h} =0$. 
Since $\Gamma_{\! q}^g (x) - J_{q}^g (x)$ commutes with itself, it thus follows that $\dot C_{q} = 0$, and that there is no associated secondary defect constraint. 

For the constraint $C_{\infty}$, the only terms in the Hamiltonian that could generate its dynamics are: 
\begin{equation}
    H' = \pi \sum_{q \in Z} \int \left(  \alpha_\zeta \bigtr{ \varphi(z)^{-2} \Gamma }{ \Gamma }{\zeta} - \beta_\zeta \bigtr{ \varphi(z)^{-2} \Delta}{\Delta}{\zeta} \right) \, dx + \frac{\pi \epsilon_\infty}{k_{\infty}} \int \killing{ \Gamma_{\infty}^{\lie f}}{\Gamma_{\infty}^{\lie f}} \, dx\ca
\end{equation} 
as the Poisson bracket of $C_{\infty}$ with all the other terms vanishes on the constraint surface. 
So that the calculation is clear, we compute the bracket of $C_{\infty}$ with each of the above terms one by one, starting with $\tr{\varphi(z)^{-2} \Gamma}{\Gamma}{\zeta}$. 
To do this, we note that: 
\begin{equation}
    \{ C_{\infty \, \textbf{1}}(x), \Gamma_{\textbf{2}}(y,z) \} = [ \mathsf{C}^{\lie h} , \Gamma_{\textbf{2}}(x,z)] \delta(x-y) - \frac{\varphi(z)}{2 \pi} \mathsf{C}_{\textbf{12}}^{\lie h} \delta'(x-y) \ca
\end{equation} 
which we use to find: 
\begin{equation}
    \pi \int \{ C_{\infty \textbf{1}} (x), \killing{\Gamma_\textbf{2}(y,z)}{\Gamma_\textbf{2}(y,z)}_{\textbf{2}} \} \, dy =- \varphi(z) \killing{\Gamma'_{\textbf{1}}(x,z)}{\mathsf{C}^{\lie h}} \ca
\end{equation} 
where the equality follows only after noting that $\Gamma(x,z)$ commutes with itself.
Using the above equation together with definition \eqref{eq:normalised-bilinear}, we find that: 
\begin{align}
    \pi \int \{ C_{\infty \textbf{1}} (x), \tr{\varphi(z)^{-2} \Gamma_\textbf{2}(y,z)}{\Gamma_\textbf{2}(y,z)}{\zeta, \textbf{2}} \} \, dy = - k_{\zeta} \, \text{Res}_{\zeta} (z-\zeta) \varphi(z)^{-1} \Gamma_{\textbf{1}}^{\lie h \, \prime} (x, z) = 0 \ca
\end{align} 
where the final equality follows from the fact that the expression in the residue has no poles at $z=\zeta$. 
This argument is also applicable to $\bigtr{ \varphi(z)^{-2} \Delta}{\Delta}{\zeta}$, and has the same result. 
To calculate the bracket of $C_{\infty}$ with the final term we recall that $\Gamma_{\infty} (x) = \sum_{q \in P_N} \Gamma_{q}(x)$, $\Delta_{\infty} (x) = \sum_{q \in P_N} \Delta_{q}(x)$ and $C_{\infty} = \Gamma_{\infty}^{\lie h} (x) - \Delta_{\infty} (x) \approx 0$. 
Thus, using the bracket: 
\begin{equation}
    \{ \Gamma_{\infty \textbf{1}} (x), \Gamma_{\infty \textbf{2}} (y) \} = -[ \mathsf{C} , \Gamma_{\infty \textbf{2}} (x)] \delta(x-y) + \frac{k_{\infty}}{2 \pi} \mathsf{C} \delta'(x-y) \ca
\end{equation}
together with the first equation in \cref{eq:project-brackets}, and the conditions $[\lie h, \lie h] \subset \lie h$, $[ \lie h , \lie f] \subset \lie f$ and $\killing{\lie f}{\lie h} = 0$ we find: 
\begin{equation}
    \{ C_{\infty \textbf{1}} (x) , \Gamma_{\infty \textbf{2}}^{\lie f} (y) = -[ C^{\lie h} , \Gamma
    _{\infty \textbf{2}}(x)]^{\lie f} \delta (x-y) = -[ C^{\lie h} , \Gamma
    _{\infty \textbf{2}}^{\lie f}(x)] \delta (x-y) \fp
\end{equation}
Hence: 
\begin{equation}
    \pi \int \{ C_{\infty \textbf{1}} (x) , \langle \Gamma_{\infty \textbf{2}}^{\lie f} (y) , \Gamma_{\infty \textbf{2}}^{\lie f} (y)\rangle = -2 \pi \langle\mathsf{C}^{\lie h},[ \Gamma_{\infty \textbf{2}}^{\lie f} (x),\Gamma_{\infty \textbf{2}}^{\lie f} (x) ] \rangle = 0 \ca
\end{equation} 
and thus there are no secondary defect constraints from $\dot C_{\infty} \approx 0$. 

\subsection{Gauge Fixing the Defect Constraints at Finite Distance \label{subsection:gauge fixing the defect constraints}} 

In this subsection, our goal is to reduce the phase space down to the constraint surface associated with the defect constraints for poles at finite distance, \textit{i.e.} $C_q \approx 0$ for all $q \in P_N$.  
During our earlier analysis of the bulk constraints, we chose to do this by introducing a series of gauge choices that render the first-class constraints second-class. 
We are not going to do this. 
Instead, we will introduce a set of gauge-invariant observables for each defect, that is, quantities which commute with the associated defect constraint. 
By working exclusively with these new variables, we can impose the constraint strongly and thus eliminate $\Delta_q (x)$. 

Each of the defects located at the poles in $P$ has an associated set of variables $(g_q, h_q, \Gamma_{\! q}, \Delta_q)$, from which we can construct three gauge-invariant observables. 
The first two of these are simply $\Gamma_q(x)$ and $\Delta_q(x)$, since these can be shown to Poisson commute with $C_q \approx 0$ using \cref{eq:gammaq-gammaq+current-bracket,eq:deltq-deltaq+current-bracket}. 
The last variable can be shown to be $\gamma_q = h_q^{-1} g_q$ from the calculation:
\begin{align}
    \{ C_{q \textbf{1}} (x) , \gamma_{q \textbf{2}} (y) \} &= \{ C_{q \textbf{1}} (x) , h^{-1}_{q \textbf{2}} (y) \} g_{q \textbf{2}} (y) + h^{-1}_{q \textbf{2}} (y) \{ C_{q \textbf{1}} (x) , g_{q \textbf{2}} (y) \} \\ 
& \hspace{-2cm}= \left( - h_{q \textbf{1}} (x) \mathsf{C} h_{q \textbf{1}}^{-1}(x) h_{q \textbf{2}}^{-1} (y) g_{q \textbf{2}} (y) + h^{-1}_{q \textbf{2}}(y) \killing{g_{q \textbf{3}}(x) g_{q \textbf{2}}(y) \mathsf{C}_{\textbf{23}} g_{q \textbf{3}}^{-1}(x)}{\mathsf{C}_{\textbf{13}}^{\lie h}} \right) \delta(x-y) = 0 \fp \nonumber
\end{align} 
The final equality in this equation is found by first integrating against a test function, after which, we use the identity $g_\textbf{1} g_\textbf{2} \mathsf{C} g^{-1}_\textbf{1} g^{-1}_\textbf{2} = \mathsf{C}$ and its equivalent for $\mathcal H$, and apply $\langle \mathsf{C}_{\textbf{23}} \, , \, \mathsf{C}_{\textbf{13}}^{\lie h} \rangle_{\textbf{3}} = \mathsf{C}_{\textbf{12}}^{\lie h}$. 
Equipped with these new variables, we can solve the constraint $C_q \approx 0$ by setting: 
\begin{equation}
    \Delta_q = \left( -\frac{k_q}{2 \pi} \gamma_q' \gamma^{-1}_q + \gamma_q \Gamma_q \gamma_q^{-1} \right)^{\lie{h}} \ca \label{eq:boundary-conditions-gamma-delta}
\end{equation} 
and thus eliminate $\Delta_q(x)$. 

Let us calculate the Poisson brackets that result from this change of variables and subsequent elimination of $\Delta_q(x)$. 
To do this, we substitute $g_q (x) = h_q (x) \gamma_q (x)$ and \cref{eq:boundary-conditions-gamma-delta} into the defect two-form $\Omega_q = \Omega_q^A + \Omega_q^B$, after which we find: 
\begin{equation}
    \Omega_q = \int \left( \frac{k_q}{4 \pi} \killing{y^{-1}_q \delta y_q}{\partial_x ( y^{-1}_q \delta y_q ) } - \killing{(y^{-1}_q \delta y_q)^2}{\Gamma_q} - \killing{y^{-1}_q \delta y_q}{\delta \Gamma_q} \right) \, dx \fp \label{eq:reduced-defect-two-form}
\end{equation} 
The simplest method of doing this calculation is to note that: 
\begin{align}
    &\killing{(h^{-1}_q \delta h_q)^2}{\Delta_q} + \killing{h^{-1}_q \delta h_q}{\delta \Delta_q} - \killing{(g^{-1}_q \delta g_q)^2}{\Gamma_q} - \killing{g^{-1}_q \delta g_q}{\delta \Gamma_q} \\
    = \, & \delta \left( \killing{ g_q^{-1} \delta g_q}{ \Gamma_q} - \killing{ h_q^{-1} \delta h_q}{ \Delta_q} \right) \nonumber \\
    = \, & \delta \left( \killing{ y_q^{-1} \delta y_q}{ \Gamma_q} + \frac{k_q}{2 \pi} \killing{h_q^{-1} \delta h_q}{y_q' y_q^{-1}}  \right) \nonumber \\
    = \, & \delta \killing{ y_q^{-1} \delta y_q}{ \Gamma_q} - \frac{k_q}{2 \pi} \killing{(h_q^{-1} \delta h_q)^2}{y'_q y^{-1}_q} - \frac{k_q}{2 \pi} \killing{y_q^{-1} h^{-1}_q \delta h_q y_q}{\partial_x (y_q^{-1} \delta y_q)} \ca \nonumber
\end{align} 
in which the final equality requires one to use the identity $\delta (y'_q y^{-1}_q ) = y_q \partial_x ( y_q^{-1} \delta y_q ) y_q^{-1}$. 
The result then follows after also using: 
\begin{align}
    \int \! dx \killing{g^{-1}_q \delta g_q}{\partial_x (g^{-1}_q \delta g_q)} = \int \! dx \left( \killing{y^{-1}_q \delta y_q}{\partial_x (y^{-1}_q \delta y_q)} \right. &+ 2 \killing{y_q^{-1} h_q^{-1} \delta h_q y_q}{\partial_x ( y^{-1}_q \delta y_q )} \nonumber \\
    + \killing{h^{-1}_q \delta h_q}{\partial_x ( h^{-1}_q \delta h_q )} &+ \left. 2 \killing{(h^{-1}_q \delta h_q)^2}{y'_q y^{-1}_q} \right) \fp 
\end{align}  
The reduced defect two-form given in \cref{eq:reduced-defect-two-form} is of the same form as $\Omega_q^A$ in \cref{eq:Initial-A-defect-two-form}, and we can thus find the Poisson brackets for $\Gamma_q(x)$ and $\gamma_q(x)$ using the methods of section \ref{subsec:gauge-fixing-step-one}. 
In particular, we find that the bracket for $\Gamma(x,z)$ with itself is unchanged, and that: 
\begin{equation}
    \{ \Gamma_{\! q \textbf{1}} (x) \, , \, \gamma_{q \textbf{2}} (y) \} = \gamma_{q \textbf{2}}(y) \mathsf{C} \delta(x-y) \fp \label{eq:Gamma-q-gamma-q-bracket}
\end{equation} 

For completeness, let us also find $\{ \Delta_{q \textbf{1}} (x) \, , \, \Delta_{q \textbf{2}} (y) \}$, $\{ \Delta_{q \textbf{1}} (x) \, , \, \Gamma_{q \textbf{2} }(y) \}$ and $\{\Delta_{q \textbf{1}} (x) \, , \,  \gamma_{q \textbf{2}} (y)\}$. 
The starting point of each of these calculations is to substitute in the expression for $\Delta_q$ given in \cref{eq:boundary-conditions-gamma-delta} and use the formulae \cref{eq:project-brackets}. 
For the first two brackets, we then make use of \cref{eq:gamma-g-J-g-bracket,eq:gamma-g-Gamma-g-bracket,eq:gammaq-gammaq+current-bracket} and project the result into $\lie h$, after which we find: 
\begin{equation}
    \{ \Delta_q (x) \, , \, \Delta_q (y) \} = \frac{k_q}{2 \pi} \, \mathsf{C} \, \delta'(x-y) - [\, \mathsf{C} \,, \Delta_{q \textbf{1}}(x)] \delta(x-y) \ca \quad \{ \Delta_q (x) \, , \, \Gamma_q(x) \} = 0 \ca
\end{equation}
as before. 
The result for the third bracket can be found by using \cref{eq:Gamma-q-gamma-q-bracket}, after which we integrate against a test function and use the identities $\gamma_{q \textbf{1}} \gamma_{q \textbf{2}} \mathsf{C} \gamma_{q \textbf{1}}^{-1} \gamma_{q \textbf{2}}^{-1} = \mathsf{C}$ and $\langle \mathsf{C}_{\textbf{23}} \, , \, \mathsf{C}_{\textbf{13}}^{\lie h} \rangle_{\textbf{3}} = \mathsf{C}_{\textbf{12}}^{\lie h}$. 
This produces: 
\begin{equation}
    \{\Delta_q (x) \, , \,  \gamma_q(y)\} = \mathsf{C}^{\lie h} \gamma_{q \textbf{2}} (y) \delta (x-y) \fp
\end{equation} 

This change of variables, and subsequent imposition of the constraints, alter both $C_{\infty} \approx 0$ and the Hamiltonian. 
The constraint $C_{\infty} \approx 0$ becomes: 
\begin{equation}
    C_{\infty} = \sum_{q \in P_N} ( \Gamma_{q} + \gamma_q \mathcal J_q \gamma_q^{-1} - \gamma_q \Gamma_q \gamma_q^{-1} )^{\lie h} \approx 0 \label{eq:gko-constraint}
\end{equation}
where we have defined $\mathcal J_q = \frac{k_q}{2 \pi} \gamma'_q \gamma_q^{-1}$, and which remains first-class. 
When $C_{\infty}$ is smeared against a test function $\mu(x,t): W \rightarrow \lie h$, it generates the infinitesimal gauge transformations of $\Gamma(x,z)$ and $\gamma_q(x)$: 
\begin{equation}
    \{ C_{\infty} [ \mu ] ,\Gamma (x,z) \} = \frac{\varphi(z)}{2 \pi} \mu'(x) + [ \Gamma(x,z) , \mu(x) ] \ca \quad \{ C_{\infty} [\mu] , \gamma_q (x) \} = \gamma_q (x) \mu (x) - \mu (x) \gamma_q (x) \fp 
\end{equation}
While within the Hamiltonian, there are two changes that occur. 
The first is that any terms that are linear in $C_q$ vanish for all $q \in P_N$, and the second is: 
\begin{align}
    \langle ( g_q (\Gamma_q - J_q) g_q^{-1})^{\lie{f}} \, , \, ( g_q ( \Gamma_q - J_q ) g_q^{-1})^{\lie{f}} \rangle = \langle ( \gamma_q ( \Gamma_q - \mathcal{J}_q ) \gamma_q^{-1})^{\lie{f}} \, , \, ( \gamma_q ( \Gamma_q - \mathcal{J}_q ) \gamma_q^{-1})^{\lie{f}} \rangle \ca
\end{align} 
which follows from $\langle X^{\lie f} \, , \, Y^{\lie f} \rangle = \langle X \, , \, Y \rangle -\langle X^{\lie h} \, , \, Y^{\lie h} \rangle$, $h \lie f h^{-1} \subset \lie f$ and $\killing{\lie f}{ \lie h} = 0$. 
We thus find the reduced Hamiltonian: 
\begin{align}
    H &= \int \killing{\mathcal N_0}{C_{\infty}} \, dx + \sum_{q \in P} \frac{\pi \epsilon_q}{k_q} \int \killing{( \gamma_q ( \Gamma_q - \mathcal{J}_q ) \gamma_q^{-1})^{\lie{f}}}{( \gamma_q ( \Gamma_q - \mathcal{J}_q ) \gamma_q^{-1})^{\lie{f}}} \\ 
    & \hspace{1.5cm} + \pi \sum_{\zeta \in Z} \int \left( \alpha_\zeta \Bigtr{ \frac{\Gamma}{\varphi(z)} }{ \frac{\Gamma}{\varphi(z)} }{\zeta} - \beta_\zeta \Bigtr{ \frac{\Lambda^{\lie h}}{\varphi(z)} }{ \frac{\Lambda^{\lie h}}{\varphi(z)} }{\zeta}  \right) \, dx \nonumber \ca
\end{align} 
where: 
\begin{equation}
    \Lambda(x,z) = \sum_{q \in P_N} \frac{\gamma_q \Gamma_q \gamma_q^{-1} - \gamma_q \mathcal J_q \gamma_q^{-1}}{z-q} \fp
\end{equation}

Let us discuss how to recover an equation for the Lax matrix of a particular $\sigma$-model using the above Hamiltonian. 
For a fixed configuration of boundary conditions, and thus boundary terms, the equations of motion for the $g_q(x)$'s produce a set of simultaneous equations. 
If the configuration is chosen carefully, these equations can be solved to find the Lax connection for a particular $\sigma$-model. 
Using the vernacular of affine Gaudin models, it is these boundary terms that determine the `realisation' of an integrable model. 

\subsection{Defining Coset Integrable Field Theories \label{section:coset-ifts}}

The constraint $C_{\infty} \approx 0$ is treated slightly differently from those that we have just discussed in that we introduce, but do not impose, a gauge fixing condition for it. 
We do this for two reasons, the first of which is simply to show that such a condition exists. 
The second reason is that it enables us to then make a comparison with the GKO construction \cite{Goddard:1984vk,Goddard:1986ee,Bais:1987zk}. 

So that this comparison is clear, let us briefly recall the construction's most rudimentary details.
The starting point is to consider a pair of (commuting) Kac-Moody currents, $J^I_+(x)$ and $J^I_{-}(x)$, at the levels $k$ and $-k$, respectively: 
\begin{equation}
    [ J_{\pm}^I (x), J^J_{\pm} (y) ] = f^{IJ}_{\hspace{3mm}K} J^{K}_{\pm} (x) \delta(x-y) \mp \frac{i k}{2 \pi} \delta^{IJ} \delta'(x-y) \ca
\end{equation} 
which are required to satisfy the first-class constraint $J_{+}^i - J_{-}^i \approx 0$. 
From these constrained currents, a two-dimensional conformal field theory $\text{Vir}[ \hat{ \lie{g} }, \hat{ \lie{h} } , k]$ can be constructed using the follow rules \cite{Bais:1987zk}: 
\begin{itemize}
    \item[$i)$] The generators of $\text{Vir}[ \hat{ \lie{g} }, \hat{ \lie{h} } , k]$ are local currents constructed by taking normal
ordered products of the currents $J_{\pm}^a (x)$ and their derivatives. 
    \item[$ii)$] $\text{Vir}[ \hat{ \lie{g} }, \hat{ \lie{h} } , k]$ contains two spin-two currents, $T(x)$ and $\bar T (x)$, whose modes each generate a copy of the Virasoro algebra, as well as higher-spin analogues that transform as Virasoro primaries. 
    \item[$iii)$] The elements of $\text{Vir}[ \hat{ \lie{g} }, \hat{ \lie{h} } , k]$ commute with the constraint $J_{+}^i - J_{-}^i \approx 0$. 
\end{itemize}
A common gauge choice used for fixing $J_{+}^i - J_{-}^i \approx 0$ is $\partial_x J_{-}^i \approx 0$, which leads to the following conditions on the modes of the currents: 
\begin{equation}
    J_{\pm}^{i \, n} \approx 0 \ca \qquad J_{+}^{i \, 0} - J_{-}^{i \, 0} \approx 0 \ca
\end{equation}
where $n$ is a non-zero integer. 
We mention this as the reader may be more familiar with the GKO construction starting from the above conditions, and because an analogous gauge choice will be introduced momentarily to fix $C_{\infty}$. 

Of course, the analogue of $J_{+}^i - J_{-}^i \approx 0$ within our construction is $C_{\infty} \approx 0$. 
With this in mind, two of the above three conditions can be translated over, allowing us to conjecture the existence of an integrable field theory, $\text{IFT}[\lie g, \lie h, \omega]$, defined by: 
\begin{itemize}
    \item[$i)$] The generators of $\text{IFT}[\lie g, \lie h, \omega]$ are local currents constructed by taking products of $\Gamma^I (x,z)$ and $\gamma_{q} (x)$ and their derivatives. 
    \item[$ii)$] The elements of $\text{IFT}[\lie g, \lie h, \omega]$ commute with the constraint $C_{\infty} \approx 0$. 
\end{itemize}
We have avoided making any concrete statement about the role of the Witt algebra (which would be the classical analogue of the Virasoro algebra in the classical context), as we are not yet completely certain of its role, especially as we expect it to be possible to construct massive integrable models. 

It is not currently possible for us to make any more general statements about the integrable hierarchies of the theories defined by the above conditions. 
This is because the latter condition implies that the resulting Poisson algebra's elements are built from $\mathcal H$-invariant polynomials about which little seems to be known. 
We expect that the best approach for determining the algebra is to use the techniques found within Drinfel'd-Sokolov reduction \cite{Drinfeld:1984qv}, and intend to discuss this in future work. 

We can fix the constraint $C_{\infty} \approx 0$ in a similar way to the approach just described if we take the gauge choice: 
\begin{equation}
    \partial_x \left( \sum_{q \in P_N}  \gamma_q \Gamma_q \gamma_q^{-1} - \gamma_q \mathcal J_q \gamma_q^{-1} \right)^{\lie h} \approx 0 \ca
\end{equation}
which due to $\mathcal B_0 = \sum_{q \in P_N} \Delta_q$ and \cref{eq:boundary-conditions-gamma-delta}, is just the Coulomb gauge $\mathcal B_0' = 0$ in disguise. 
This can be done for the following reason (see \cite{Bowcock:1988xr} for an earlier application of this argument in the context of the Gauged Wess-Zumino-Witten model). 
Earlier on, we noted that $\mathcal N_0 (x,t)$ arises as a constant of integration, which we are free to set to zero. 
After making this choice, it is still possible to make $t$-independent gauge transformations, as these preserve the condition $\mathcal N_0 (x,t) = 0$. 
Using such a transformation, we can set $\mathcal{B}_0'(x,0) = 0$. 
Once $\mathcal N_0  (x,t) = 0$ is plugged into the equation of motion $\dot{\mathcal{B}}_0 = \mathcal N_0' + [\mathcal B_0, \mathcal N_0]$, we find that $\dot{\mathcal B}_0 = 0$, from which the Coulomb gauge condition then follows.

\section{Conclusions and Future Directions \label{sec:future-directions}} 

In this article, we have performed a Hamiltonian analysis of doubled four-dimensional Chern-Simons theory by following Dirac's algorithm. 
During the analysis, we have also developed a constraint characterisation of the theory's boundary conditions and analysed them using Dirac's methods. 
This enabled us to impose all but one of the boundary constraints by introducing a set of gauge-invariant observables. 
The resulting Poisson algebra is that of an affine Gaudin model subject to the constraint \cref{eq:gko-constraint}, generalising the GKO construction to the world of integrable models. 

Let us conclude by discussing several potential generalisations of the above work. 
Throughout this paper, two assumptions have greatly simplified our analysis. 
These were: $i)$ the requirement that our field configurations satisfy gauged chiral/anti-chiral boundary conditions, and $ii)$ the restriction imposed upon $\omega$ that it has at most first-order zeros and simple poles. 

From the Lagrangian analysis, we know that a sufficient condition for the admissibility of a given boundary condition is that it satisfies $\killing{A^{\lie f}}{\delta A^{\lie{f}}} = 0$. 
This equation has many possible solutions, allowing us to relax assumption $i)$ and use a different condition. 
When this is done, the above analysis needs to be modified by adding different improvement terms to the action and Gauss law constraints, ensuring they remain functionally differentiable. 
Of course, this change will modify the Hamiltonian, and thus could affect whether a set of secondary constraints appears during the analysis of the boundary constraints. 
While we do not expect this to be the case, we do not currently have proof of it. 

There are two different ways to relax assumption $ii)$, where the first is to allow $\omega$ to have higher-order zeros, and the second, higher degree poles. 
In the former case,  the main change to our analysis will be that the Lax matrices $\mathcal A$ and $\mathcal B$ can have higher-order poles. 
Depending on the boundary conditions that one chooses to impose at the points in $Z$, this will then alter $\mathcal M$, $\mathcal N$ and the associated terms in the Hamiltonian. 
We do not expect this to result in any changes to the Poisson algebra for $\Gamma(x,z)$ and $\{g_q (x)\}$, or to the constraint $C_{\infty} \approx 0$. 

Conversely, we expect that allowing for $\omega$ to have higher degree poles will radically change both the Poisson algebra and the constraints. 
We believe that the correct approach to this analysis lies in the regularisation procedure introduced in \cite{Benini:2020skc}. 
Working within this framework, we expect that each pair $(\Gamma_q,\gamma_q)$ is extended such that $\Gamma_q$ is replaced by a set of Takiff currents $\{ \Gamma_q^{[a]}\}$, where $a = 0, \ldots , -1+\text{deg} \, p $, and by $\{ \partial^a \gamma_q \}$. 
Further still, we expect that the boundary condition $A^{\lie{h}}_{W}(q) = B_W (q)$ (which eventually became the boundary constraint) is replaced by: 
\begin{equation}
    \partial^a A^{\lie h}_W (q) - \partial^a B_W(q) = 0 \ca
\end{equation} 
and thus that there will be new conditions to supplement $C_{\infty} \approx 0$. 
We intend to discuss this elsewhere. 

In addition to these generalisations, the above analysis also opens up six other potential avenues of research. 
The first of these is the potential existence of extended quantum groups. 
To illustrate what is meant by this, we recall a few basic facts from the doubling of three-dimensional Chern-Simons given in \cite{Moore:1989yh}, in which a $\mathcal G$-theory was coupled to an $\mathcal H$-theory at a boundary. 
For exactly the reason given in section \ref{section:common-centre}, this coupling produces a theory with the gauge group $\mathcal{G \times H / Z}$. 
This quotient affects the Wilson line spectrum within the resulting theory in three different ways: $a)$ those Wilson lines that are in a representation of $\mathcal{G \times H}$ but not $\mathcal{G \times H / Z}$, act trivially; $b)$ Wilson lines that are equivalent under a `spectral-flow' operation are identified; and $c)$ there exist new operators in the $\mathcal{G \times H / Z}$ theory which cannot be decomposed into its Wilson lines, but that are given by a composition of lines in the $\mathcal{G \times H}$ theory. 
Given that the fusion algebra for Wilson lines within 4dCS reproduces the structure of quantum groups for an appropriate replacement of $\mathbb C P^1$ and $\omega$, we expect to find similar extensions from doubled 4dCS. 

The second avenue concerns the doubling of six-dimensional Chern-Simons on twistor space  (6dCS) given in \cite{Cole:2024sje}, within which a range of integrable deformations were constructed. 
There are two directions that we believe this work could be taken in. 
The first is to perform the Hamiltonian analysis of the theory, which we believe can be possible using the methods developed above. 
The second direction, which is contingent on the first, is to study the quantisation of the resulting Poisson algebra. 
The hope is that the resulting structure somewhat resembles the GKO construction in two-dimensions, and that we can thus find a plethora of solvable quantum field theories in both three and four-dimensions via symmetry reduction of 6dCS. 

Within the theory of classical $\mathcal W$-algebras there exists the affine Harish-Chandra isomorphism: 
\begin{equation}
    \mathcal W (\lie g^{\vee}) \cong \lie z (\hat{\lie g}) \fp
\end{equation} 
The object on the left-hand side of this equivalence is a $\mathcal W$-algebra that is constructed via the Drinfel'd-Sokolov reduction of a Poisson vertex algebra $\mathcal V (\lie g^\vee)$, where $\lie g^\vee$ is the Langlands dual of $\lie g$. 
While the object on the right-hand side is the Feigin-Frenkel centre of the affine vertex algebra at critical level $V_{-h^\vee} (\lie g)$ equipped with a particular Poisson bracket. 
The reader can find further information about both of these objects and the isomorphism in the book \cite{Molev}. 
Suppose that we fix $\omega = k \, dz/z$, in both a normal and doubled version of 4dCS. 
In the former case, this is known to reduce to the Wess-Zumino-Witten model \cite{Costello:2019tri}, while the latter simply becomes the (classical) GKO construction. 
Thus, we expect that the above isomorphism can be reformulated as an equivalence between the classical doubled 4dCS theory with gauge group\footnote{The superscript on these groups indicates the level in the Poisson algebra of that particular theory.} $\mathcal G^{(k)} \times \mathcal G^{(1)} \times \mathcal G_{\text{diag}}^{(k+1)} / \mathcal Z$ and the quantisation of the normal 4dCS $\mathcal G^{(k)}$-theory equipped with a particular Poisson bracket. 
Our third question is thus: can this construction be generalised? And if so for what choices of $\mathcal G$, $\mathcal H$ and $\omega$ does there exists a `doubly-affine' Harish-Chandra Isomorphism:
\begin{equation}
    \mathcal T_{D} (\mathcal{G,H},\omega) \cong \mathcal T_0 (\mathcal{G},\omega) \ca
\end{equation} 
where on the left-hand side we have the doubled 4dCS theory, and on the right-hand side, some sub-sector of the standard one. 

The final three questions that we think would be interesting are as follows. Firstly, what are the Poisson algebras for the (constrained) finite-dimensional models that arise from the circle reduction of doubled 4dCS, and are they interesting? Secondly, what is the relationship between the Poisson algebra defined by the conditions given in section \ref{section:coset-ifts} and Kac et al's recent work on Poisson vertex algebras, see \textit{e.g.} the review \cite{desole2023poissonvertexalgebrashamiltonian}?
And finally, is there a soft theorem associated with the improper transformations discussed in \ref{sec:edgemodes}, and if so, what does it tell us about integrable models? 

\appendix

\section{Boundary Integrals\label{appendix:Boundary-Integrals}} 

In this appendix, we present a variation on Lemma 2.2 of \cite{Benini:2020skc}, which was inspired by Remark 2.1 of \cite{Vicedo:2019dej}. 
Our goal is to, integrate over $\mathbb C P^1_{\omega}$ in: 
\begin{equation}
    I = \frac{i}{8 \pi^2} \int_M d(\omega \wedge \eta) \fp
\end{equation}
where $\eta \in \Omega^2(M)$ satisfies the assumptions given in section \ref{sec: Action}. 

We begin this calculation by thickening each puncture $p \in P \cup Z$ to a small disc $B_p$ of radius $r_p < R_p$, where these radii are chosen to ensure the discs remain disjoint. 
Doing this introduces a boundary $\partial M$ into $M$ given by $\partial M = \sqcup_{p \in P \cup Z} ( \partial B_p \times W)$. 
For each of the (disjoint) components $\partial M_p := \partial B_P \times W$ of $\partial M$ there is an associated inclusion:
\begin{equation}
    \iota_p : \partial B_p \times W \rightarrow M \ca
\end{equation} 
that embeds $\partial B_p \times W$ into $M$. 
Hence, by Stokes' theorem, we have:
\begin{equation}
    I = \sum_{p \in P \cup Z} I_p \ca \quad \text{where} \quad I_p =\frac{i}{8 \pi^2} \int_{\partial M_p} \iota_p^* \omega \wedge \iota_p^* \eta \ca
\end{equation} 
in which $\iota_p^*$ is the pullback of $\iota_p$. 
The integrals $I_p$ are either associated with $a)$ a pole $q \in P$, or $b)$ a zero $\zeta \in Z$. 
We will therefore deal with these two cases separately. 

For case $a)$ with $q \in P_0$ we start by introducing a local set of polar coordinates defined by $z = q + r \exp i \theta$ in terms of which $I_q$ is\footnote{The coefficient of this integral is positive because $\partial M_q$ is a lower bound and so introduces an additional minus sign that cancels $i^2=-1$.}: 
\begin{equation}
    I_q = \frac{1}{8 \pi^2} \int_{\partial M_q} d \theta \wedge \left( k_q \eta(r_q,\theta) + \sum_{q' \in P_0\setminus q} k_{q'} \frac{e^{i \theta}}{e^{i \theta} +\tfrac{q-q'}{r_q}} \eta (r_q,\theta) \right) 
\end{equation}
Due to assumption $1.$ $\eta(r_q,\theta)$ has a Fourier expansion $\eta(r_q,\theta) = \sum_{n = \infty}^{\infty} \eta_n(r_q) \exp (i n \theta)$ in which all coefficients other than $\eta_0$ vanish in the limit $r_q \rightarrow 0$. 
Using this series, we can integrate over $\theta$ to find: 
\begin{align}
    I_q &=\frac{1}{8 \pi^2} \int_{W_q} \left( k_q \eta_0(r_q) + \sum_{q' \in P_0\setminus q} k_{q'} \sum_{n=-\infty}^{\infty} \frac{e^{i (n+1) \theta}}{e^{i \theta} +\tfrac{q-q'}{r_q}}  \eta_{n} (r_q) \right) \nonumber \\
    &=\frac{k_q}{4 \pi} \int_{W_q} \eta_0(r_q) + \frac{1}{4 \pi}\sum_{q' \in P_0\setminus q} k_{q'} \sum_{n=1}^{\infty} (-1)^{1-n} \left( \frac{r_q}{q-q'} \right)^n \int_{W_q} \eta_{-n} (r_q) \ca \label{eq:Iq-1} 
\end{align} 
where in the second equality we used:
\begin{equation}
    \int_0^{2 \pi}d \theta \, \frac{e^{i (n+1) \theta}}{e^{i \theta} + a} = \begin{cases}
        (-1)^{n+1} 2 \pi a^n & \text{for} \ \  n<0 \ca \\
        0 & \text{otherwise} \ca 
    \end{cases} \label{eq:angular-integral-identity}
\end{equation}
 $|a|>1$, which is true for us because $|q-q'| > r_q$ by construction. 
Thus, after taking the limit $r_q \rightarrow 0$, which shrinks the disc back to a point, we find: 
\begin{align}
    I_q = \frac{k_q}{4 \pi} \int_{W_q} \iota_q^* \eta \ca
\end{align}
where we have abused notation to use $\iota_q^* \eta = \eta(q) = \eta_{0}(0)$. 
A similar calculation can be performed for the pole at infinity after introducing polar coordinates defined by $z^{-1} =r\exp i \theta$. 
This yields: 
\begin{equation}
    I_{\infty} = \frac{k_{\infty}}{4 \pi^2} \int_{W_{\infty}} \iota_\infty^* \eta \fp
\end{equation} 

Let us turn now to case $b)$. 
We start by substituting in $\eta = (z-\zeta)^{-m} \eta^{(\zeta)}$, as in assumption $2)$, and then introduce a set of polar coordinates defined by $z=\zeta+r\exp i \theta$, after which we find: 
\begin{equation}
    I_\zeta = \frac{1}{8 \pi^2} \int_{\partial M_\zeta} d \theta \wedge \left( \frac{1}{r_\zeta^{m}} \sum_{q \in P} k_q  \frac{e^{i (1-m) \theta} }{e^{i \theta} + \tfrac{\zeta - q}{r_\zeta }} \right) \eta^{(\zeta)}(r_\zeta,\theta) \fp
\end{equation} 
The next step in this calculation is to expand $\eta^{(\zeta)} (r_\zeta,\theta)$ into a Fourier series, which we can do because of assumption $2)$, and evaluate the integral over $\theta$. 
This calculation is substantively the same as the one performed in \cref{eq:Iq-1}, and required the use of \cref{eq:angular-integral-identity}, which holds because $|\zeta - q | > r$, again by construction. 
Thus, we find: 
\begin{align}
    I_\zeta &= \frac{1}{8 \pi^2} \int_{\partial M_\zeta} d \theta \wedge \left( \frac{1}{r_\zeta^{m}} \sum_{q \in P} k_q \sum_{n=-\infty}^\infty \frac{e^{i (1+n) \theta} }{e^{i \theta} + \tfrac{\zeta - q}{r_\zeta }} \eta^{(\zeta)}_{n+m}(r_\zeta) \right) \nonumber \\
    &= \frac{1}{4 \pi} \sum_{q \in P} k_q  \sum_{n=1}^{\infty} \frac{(-1)^{n+1}}{r_\zeta^{m}} \left(  \frac{r_\zeta}{\zeta-q} \right)^n \int_{W_\zeta} \eta^{(\zeta)}_{m-n}(r_\zeta) \nonumber \\
    &= \frac{r_\zeta^{1-m}}{4 \pi}  \left( \sum_q \frac{k_q}{\zeta-q}\right) \int_{W_\zeta} \eta_{m-1}^{(\zeta)} (r_{\zeta}) + \frac{r_\zeta^{2-m}}{4 \pi}  \left( \sum_q \frac{-k_q}{(\zeta-q)^2}\right) \int_{W_\zeta} \eta_{m-2}^{(\zeta)} (r_{\zeta}) + \cdots \nonumber \\
    &=\frac{r_\zeta^{2-m}}{4 \pi} k_\zeta \int_{W_\zeta} \eta_{m-2}^{(\zeta)} (r_{\zeta}) + \frac{1}{4 \pi} \sum_{q \in P} k_q  \sum_{n=3}^{\infty} \frac{(-1)^{n+1}}{r_\zeta^{m}} \left(  \frac{r_\zeta}{\zeta-q} \right)^n \int_{W_\zeta} \eta^{(\zeta)}_{m-n}(r_\zeta) \ca
\end{align} 
where in the final equality we have used $\varphi(\zeta) = \sum_q \tfrac{k_q}{\zeta-q} = 0$, as $\zeta$ is a zero of $\omega$, and set $k_\zeta=\varphi'(\zeta) = - \sum_q \tfrac{k_q}{(\zeta -q )^2}$, which is easy to verify. 
To finish, we shrink the disc $B_\zeta$ to a point by taking the limit $r_{\zeta} \rightarrow$ and find $I_{\zeta} = 0$, if $m=0,1$, or: 
\begin{equation}
    I_{\zeta} = \frac{k_\zeta}{4 \pi} \int_{W_\zeta} \iota_p^* \eta^{(\zeta)}
\end{equation}
when $m=2$; again we have abused notation by setting $\iota_p^* \eta^{(\zeta)} = \eta^{(\zeta)}(\zeta) = \eta^{(\zeta)}(0)$. 

To summarise, by cutting out small discs containing the locations of $\omega$'s poles and zeros, calculating the integral, and then shrinking the discs to a point, we have found: 
\begin{equation}
    I = \sum_{q \in P} \frac{k_q}{4 \pi} \int_{W_q} \iota^*_q \eta + \sum_{\zeta \in Z} \frac{k_\zeta}{4 \pi} \int_{W_q} \iota^*_\zeta \eta^{(\zeta)} \ca
\end{equation} 
where $\iota^*_\zeta \eta^{(\zeta)} = \res{\zeta} (z-\zeta) \, \eta$ vanishes unless $\eta$ has a second degree pole at $\zeta$. 

\printbibliography

\end{document}